# Glueballs and strings in $Sp(2N)$ Yang-Mills theories

Ed Bennett[1,*] Jack Holligan,[2,3,†] Deog Ki Hong[4,‡] Jong-Wan Lee,[4,5,§] C.-J. David Lin,[6,7,∥] Biagio Lucini[1,8,¶]
Maurizio Piai,[2,**] and Davide Vadacchino[9,10,††]

[1]*Swansea Academy of Advanced Computing, Swansea University (Bay Campus),
Fabian Way, SA1 8EN Swansea, Wales, United Kingdom*
[2]*Department of Physics, College of Science, Swansea University (Park Campus),
Singleton Park, SA2 8PP Swansea, Wales, United Kingdom*
[3]*The Institute for Computational Cosmology (ICC), Department of Physics, South Road,
Durham DH1 3LE, United Kingdom*
[4]*Department of Physics, Pusan National University, Busan 46241, Korea*
[5]*Extreme Physics Institute, Pusan National University, Busan 46241, Korea*
[6]*Institute of Physics, National Chiao-Tung University, 1001 Ta-Hsueh Road, Hsinchu 30010, Taiwan*
[7]*Centre for High Energy Physics, Chung-Yuan Christian University, Chung-Li 32023, Taiwan*
[8]*Department of Mathematics, College of Science, Swansea University (Bay Campus),
Fabian Way, SA1 8EN Swansea, Wales, United Kingdom*
[9]*INFN, Sezione di Pisa, Largo Pontecorvo 3, 56127 Pisa, Italy*
[10]*School of Mathematics and Hamilton Mathematics Institute,
Trinity College, D02 PN40 Dublin 2, Ireland*

Motivated in part by the pseudo-Nambu Goldstone boson mechanism of electroweak symmetry breaking in composite Higgs models, in part by dark matter scenarios with strongly coupled origin, as well as by general theoretical considerations related to the large-$N$ extrapolation, we perform lattice studies of the Yang-Mills theories with $Sp(2N)$ gauge groups. We measure the string tension and the mass spectrum of glueballs, extracted from appropriate two-point correlation functions of operators organized as irreducible representations of the octahedral symmetry group. We perform the continuum extrapolation and study the magnitude of finite-size effects, showing that they are negligible in our calculation. We present new numerical results for $N = 1, 2, 3, 4$, combine them with data previously obtained for $N = 2$, and extrapolate toward $N \to \infty$. We confirm explicitly the expectation that, as already known for $N = 1, 2$ also for $N = 3, 4$ a confining potential rising linearly with the distance binds a static quark to its antiquark. We compare our results to the existing literature on other gauge groups, with particular attention devoted to the large-$N$ limit. We find agreement with the known values of the mass of the $0^{++}$, $0^{++*}$, and $2^{++}$ glueballs obtained taking the large-$N$ limit in the $SU(N)$ groups. In addition, we determine for the first time the mass of some heavier glueball states at finite $N$ in $Sp(2N)$ and extrapolate the results toward $N \to +\infty$ taking the limit in the latter groups. Since the large-$N$ limit of $Sp(2N)$ is the same as in $SU(N)$, our results are relevant also for the study of QCD-like theories.

_________________
[*]e.j.bennett@swansea.ac.uk
[†]J.HOLLIGAN.968137@swansea.ac.uk
[‡]dkhong@pusan.ac.kr
[§]jwlee823@pusan.ac.kr
[∥]dlin@mail.nctu.edu.tw
[¶]b.lucini@swansea.ac.uk
[**]m.piai@swansea.ac.uk
[††]vadacchd@tcd.ie

## I. INTRODUCTION

Recent years have seen a resurgence of interest in gauge theories based on symplectic groups, driven by theoretical as well as phenomenological motivations, related to model building in the context of physics beyond the Standard Model (SM). In comparison with the $SU(N)$ and [limited to $(2 + 1)$-dimensional] $SO(N)$ cases [1–5], the literature on lattice studies of Yang-Mills theories with $Sp(2N)$ gauge groups is limited in its extent, scope, and reach (see, for instance, Ref. [6]). In a recent publication [7] (see also Refs. [8–11]), some of us announced the intention to carry out a long-term, systematic lattice exploration of the strong-coupling dynamics of the theories based on $Sp(2N)$ gauge symmetry, and proposed a research program that includes as one of its crucial steps the study of the dynamics of glueballs and strings in the pure gauge theory.

In the same paper [7], we presented our first comprehensive study of the $Sp(4)$ pure gauge theory, and computed the spectrum of masses and decay constants of mesons consisting of two fundamental (Dirac) fermions, treated in the quenched approximation. In the context of composite Higgs models (CHMs) [12–14] (see also Refs. [15–55]), that initial step provided an important source of quantitative information about the underlying dynamics. In particular, we started to explore and exploit the dynamical origin of low-energy effective field theories (EFT) based upon the $SU(4)/Sp(4)$ coset, which have a prominent role in the CHM context (see, for instance, Refs. [56–76]), as well as for related models of dark matter with a strong-coupling origin [77–80]. More recently [81], some of us presented the first continuum results of the lattice study of the $Sp(4)$ theory with dynamical Wilson fermions, hence making the treatment of the dynamics more realistic and useful in the CHM context. A first set of exploratory studies of the quenched theory with valence fermions in multiple representations has been published in Ref. [82].

In the present paper (see also Refs. [83,84]), we take major steps in a complementary direction, by focusing on the pure gauge theory without matter content, but extending the analysis to different $Sp(2N)$ gauge groups. Our specific objective is to obtain for the $Sp(2N)$ Yang-Mills theories in $D = 3 + 1$ dimensions a comparable level of control over the spectra of strings and glueballs as achieved for the previously studied $SU(N)$ and $SO(N)$ gauge theories [1–5]. On a theoretical side, this endeavor will allow us to study the approach toward the common large-$N$ limit via an alternative sequence of groups in respect to $SU(N)$ and $SO(N)$. In turn, this will provide an alternative set of numerical tests for such conjectural behaviors as those put forward, for example, in Refs. [83,85–92], as well as allowing comparison to calculations performed within the context of gauge-gravity dualities (see, for instance, Refs. [93–103]) or with alternative field theoretical methods [104–108]. In pragmatic terms, we will also set the stage for future studies in quenched theories realizing the $SU(4)/Sp(4)$ coset, based upon generic $Sp(2N)$ groups. In [82] some of us studied the quenched meson spectrum of a theory with $Sp(4)$ gauge group and fermions in multiple representations, relevant for the implementation of partially composite top scenarios. The present study is a first step in the direction of extending these results to the determination of the dependence on $N$ of the masses and decay constants of the states of mesonic spectrum for $Sp(2N)$ gauge theories, both in the fundamental and in higher dimensional representations of the gauge group.

In our investigation, we adopt a unified approach to the study of $Sp(2N)$ gauge theories, by applying the same heat bath (HB) algorithm exploited in Ref. [7] for the $Sp(4)$ theory to the whole $Sp(2N)$ sequence. In addition to reconsidering $Sp(2) \sim SU(2)$, which allows us to test our algorithm and procedures by comparing to existing results in the literature and extending the $N = 2$ results discussed in Ref. [7] with new calculations, we consider the $N > 2$ cases. For the latter, with the exception of our study in Ref. [83] (focusing on a discussion of the two lowest-lying glueball states and on a remarkable universality property of their ratio) and Ref. [84] (presenting some preliminary numerical results, further discussed in the current work), no detailed calculation of the glueballs has been reported in the literature so far. From an operational perspective, we first compute the effective string tension and glueball masses in the large-volume limit for fixed lattice spacing and $N = 1, 2, 3, 4$. Then, after taking the continuum limit of the glueball spectrum at each investigated value of $N$, we perform a critical analysis of the large-$N$ extrapolation and compare to other results in the literature, as appropriate.

The paper is organized as follows. In Sec. II we introduce the basic definitions and conventions adopted in the lattice calculations. In Sec. III we describe the spectral observables of interest. In Sec. IV we present our numerical results, including also the extrapolations to continuum and large-$N$ limits. Section V summarizes our conclusions and suggestions for future further inquiries. We have relegated some important technical details to the appendixes.

## II. NUMERICAL SIMULATIONS OF THE LATTICE MODEL

In four Euclidean dimensions, the $Sp(2N)$ gauge theory is defined by the following action:

$$S_{YM} \equiv -\frac{1}{2g_0^2} \int d^4x \, \mathrm{Tr} F_{\mu\nu} F_{\mu\nu}, \qquad (1)$$

where $g_0$ is the gauge coupling, the trace is over color indices, the field-strength tensor $F_{\mu\nu} \equiv \sum_A F_{\mu\nu}^A \tau^A$ is defined by

$$F_{\mu\nu}^A \equiv \partial_\mu A_\nu^A - \partial_\nu A_\mu^A + f^{ABC} A_\mu^B A_\nu^C, \qquad (2)$$

and the gauge fields are $A_\mu = \sum_A A_\mu^A \tau^A$, with the indices taking the values $A, B, C = 1, \ldots, N(2N+1)$, for $Sp(2N)$. The $2N \times 2N$ matrices $\tau^A$ are the generators of the algebra associated with the $Sp(2N)$ group, written in the fundamental representation, and normalized according to $\mathrm{Tr}\tau^A\tau^B = \frac{1}{2}\delta^{AB}$. The structure constants of the algebra are defined as the commutation relations

$$[\tau^A, \tau^B] = i f^{ABC} \tau^C. \qquad (3)$$

We regularize the theory on a lattice, in which the continuum coordinates are discretized with lattice spacing $a$.

The four-dimensional Euclidean hypercubic lattice consists of sites that are denoted by their position $x$ in the lattice. The sites are connected by links that are characterized by the position $x$ and direction $\mu$, where $\mu, \nu = 0, \ldots, 3$ label the four spacetime coordinates. The elementary variables of the lattice regularized $Sp(2N)$ gauge theory are the *link variables*, defined as

$$U_\mu(x) \equiv \exp\left( i \int_x^{x+\hat{\mu}} d\lambda^\mu \tau^A A_\mu^A(\lambda) \right), \qquad (4)$$

with $\hat{\mu}$ the unit vector in direction $\mu$. The $2N \times 2N$ matrices $U_\mu(x)$ transform according to the fundamental representation of the $Sp(2N)$ group. Gauge transformations take the form $U_\mu(x) \to g(x) U_\mu(x) g^\dagger(x+\hat{\mu})$, with $g(x)$ a group element.

The simplest gauge invariant operator is the trace of the product of link variables around an elementary square of the lattice,

$$\mathcal{P}_{\mu\nu}(x) \equiv U_\mu(x) U_\nu(x+\hat{\mu}) U_\mu^\dagger(x+\hat{\nu}) U_\nu^\dagger(x). \qquad (5)$$

The matrices $\mathcal{P}_{\mu\nu}(x)$ are called the *elementary plaquette variables* or just *plaquettes* for short.

The $Sp(2N)$ lattice gauge theory (LGT) we adopt in this paper is defined by the *Wilson action*,

$$S_W \equiv \beta \sum_x \sum_{\mu < \nu} \left( 1 - \frac{1}{2N} \Re \mathrm{Tr} \mathcal{P}_{\mu\nu}(x) \right). \qquad (6)$$

In this expression, $\Re \mathrm{Tr} \mathcal{P}_{\mu\nu}(x)$ is the real part of the trace of $\mathcal{P}_{\mu\nu}(x)$. The *inverse coupling* $\beta$ is related to $g_0$ by the request that, when the lattice spacing $a \to 0$, Eq. (6) tends to the continuum Yang-Mills action in Eq. (1), at leading order in $a$. From this requirement, one finds

$$\beta = \frac{4N}{g_0^2}. \qquad (7)$$

Monte Carlo numerical evaluations of the integrals appearing in the definitions allow us to explore the long-distance regime of the $Sp(2N)$ (pure) Yang-Mills theories, capturing nonperturbative phenomena that are not accessible to perturbation theory. For any quantity $\mathcal{O}(A_\mu)$ that depends on the gauge fields, the physical observables are estimated as ensemble averages, which are schematically given by

$$\langle \mathcal{O}(U_\mu) \rangle \equiv \frac{\int \mathcal{D} U_\mu e^{-S_W} \mathcal{O}(U_\mu)}{Z(\beta)}, \qquad (8)$$

where the denominator is

$$Z(\beta) \equiv \int \mathcal{D} U_\mu e^{-S_W}. \qquad (9)$$

These expressions can be computed numerically by sampling the space of configurations of $U_\mu(x)$, according to the probability distribution $e^{-S_W}$. This can be achieved by defining a Markovian process that evolves a particular configuration according to an *update algorithm*. The algorithm must respect detailed balance and reproduce the correct equilibrium distribution. Then, if $i$ labels the $M$ configurations produced sequentially, the ensemble average can be obtained as the simple average

$$\langle \mathcal{O} \rangle = \lim_{M \to \infty} \frac{1}{M} \sum_{i=1}^M O_i, \qquad (10)$$

where $O_i$ is the value that the observable $\mathcal{O}(U_\mu)$ takes on configuration $i$. The algorithm adopted in this work to produce successive configurations is a combination of local HB and overrelaxation (OR) updates, adapted to $Sp(2N)$ from the $SU(2N)$ implementation provided in Ref. [109] (see Appendix A for further details). Configurations are updated sequentially, one link at a time, with one HB update followed by four OR updates. An update of all the links on the lattice is called a *lattice sweep*. Successive configurations produced in this manner are correlated; to reduce the effects of autocorrelation, the ensemble averages used for physical calculations are restricted by sampling the history in steps that are separated by ten lattice sweeps. Our implementation of the algorithms above is based on the HIREP code [110], originally designed for the treatment of $SU(N)$ theories with matter fields in general representations.[1]

The lattice size being finite, we impose periodic boundary conditions in all directions. In the continuum, it is known that resulting configurations of gauge fields are characterized by an integer topological number [111], defined as

$$Q \equiv \frac{1}{32\pi^2} \epsilon_{\mu\nu\rho\sigma} \int d^4 x \mathrm{Tr} F_{\mu\nu} F_{\rho\sigma}. \qquad (11)$$

The associated susceptibility can be related to the large mass of the $\eta'$ particle [112]. The configuration space is thus divided into sectors, each characterized by an integer value of the topological number $Q$, and separated from each other by potential barriers.

Because of the lattice discretization, the topological charge $Q$ takes *nearly* integer values [113–115]. There are many microscopic lattice definitions of the topological charge that reproduce the same, correct long-distance results in the $a \to 0$ limit. In this work we adopt the definition

---

[1]HIREP can be downloaded from https://github.com/claudiopica/HiRep.

$$Q \equiv \sum_x q(x), \tag{12}$$

with

$$q(x) \equiv \frac{1}{32\pi^2} \epsilon_{\mu\nu\rho\sigma} \mathrm{Tr}\{U_{\mu\nu}(x)U_{\rho\sigma}(x)\}, \tag{13}$$

and where $x$ runs upon all lattice sites. Since these definitions make use of the short-distance degrees of freedom, calculations are affected by short-range fluctuations. These effects can be reduced by the use of smoothing operations such as the gradient (or Wilson) flow [116], which we will introduce below.

As in the continuum, also on the lattice the different topological sectors are separated by potential barriers. If these barriers are not too steep, in simulations a sufficient number of tunneling events between sectors will occur, and the resulting measured topological charge will be Gaussian distributed around zero. However, superselection of topological sectors can be shown to emerge close to the continuum limit [113,114]. As a consequence, Monte Carlo update algorithms tend to become trapped inside one of the topological sectors. Hence, close to the continuum limit, the topological charge has a long autocorrelation time. This phenomenon is referred to in the literature as *topological freezing*. Because of large-$N$ suppression of small-size instantons, which are crucial for changing the topological charge in numerical simulations [117], topological freezing becomes more severe as $N$ increases. We shall discuss implications of this algorithmical trapping more extensively later in the paper, focusing on the effects of topological freezing on the observables that are of interest to us.

To remove ultraviolet fluctuations that would otherwise dominate the signal in the extraction of the topological charge, we employ the gradient flow [116,118] of the Wilson action (i.e., the Wilson flow). The gradient flow provides a first-principles approach to the smoothening of configurations with efficiency comparable to that of the more empirical and time-honored cooling methods (see, for instance, Ref. [119]). Moreover, the evolution of observables under the gradient flow can be determined with numerical procedures that can easily be implemented. For this reason, this method has gained a prominent role in lattice studies in recent years.

With $\mathfrak{t}$ the coordinate in an additional fifth dimension (referred to as *flow time*) and $x$ a point in four-dimensional space, the gradient flow $B_\mu(\mathfrak{t}, x)$ is defined by the following differential equations and boundary conditions:

$$\frac{dB_\mu(\mathfrak{t}, x)}{d\mathfrak{t}} = D_\nu G_{\nu\mu}(\mathfrak{t}, x), \quad \text{with} \quad B_\mu(\mathfrak{t} = 0, x) = A_\mu(x). \tag{14}$$

Here $A_\mu(x)$ is the continuum gauge field, while the covariant derivative is $D_\mu = \partial_\mu + [B_\mu, \cdot]$, which yields the field-strength tensor:

$$G_{\mu\nu} \equiv [D_\mu, D_\nu]. \tag{15}$$

On the lattice, the gradient flow for the action in Eq. (6) is defined by

$$\frac{\partial V_\mu(\mathfrak{t}, x)}{\partial \mathfrak{t}} \equiv -g_0^2 \{\partial_{x,\mu} S^{\text{flow}}[V_\mu]\} V_\mu(\mathfrak{t}, x), \tag{16}$$

with initial condition $V_\mu(\mathfrak{t} = 0, x) = U_\mu(x)$. Here, $S^{\text{flow}}$ is the Wilson plaquette action for $V_\mu$.

The gradient flow describes a diffusion process with time $\mathfrak{t}$. At the leading order in the coupling $g_0$, the flow to time $\mathfrak{t}$ acts on the gauge fields as a Gaussian spherical smoothing operation, with root-mean-square radius $\sqrt{8\mathfrak{t}}$, the flow time $\mathfrak{t}$ having the dimension of a length squared. Furthermore, to all orders in perturbations in $g_0$, any gauge invariant composite operator constructed from $B_\mu(\mathfrak{t}, x)$ is renormalized at $\mathfrak{t} > 0$, and thus directly encodes physically observable properties. Using a value of the flow time $\tau$ such that $a \ll \sqrt{8\tau} \ll R$, where $R$ is a typical hadronic scale, provides four-dimensional smoothed configurations $V(\tau, x)$ that are not affected by ultraviolet fluctuations and still encode the correct infrared behavior of the theory.

## III. THE SPECTRUM

Non-Abelian Yang-Mills theories confine, and their spectra consist of massive color-neutral states called *glueballs*. If a non-Abelian gauge theory is formulated on a space with one or more compact directions, wrapping *torelon* states arise. The validity of the confinement picture for the specific case of $Sp(4)$ has been confirmed explicitly in the numerical calculations reported in Refs. [6,7]. The main objectives of this work are to show through lattice calculations that, as one would expect, confinement arises also in $Sp(6)$ and $Sp(8)$, to measure the resulting glueball mass spectrum, and to determine the large-$N$ limit of the latter. Before discussing our numerical results, in this section we review the methodology we shall adopt. The methodological material presented in this section is based upon notions that have been tested and are well established in the literature. Details beyond our exposition can be found, e.g., in Refs. [120–125], from which we draw heavily in what follows.

### A. The variational method

Let $\mathcal{H}$ be a Hamiltonian of the three-dimensional system of volume $L^3$ defined on a lattice with $L_t$ time slices.[2] Let $|n\rangle$ and $E_n$ be the eigenstates and eigenvalues of $\mathcal{H}$, i.e.,

$$\mathcal{H}|n\rangle = E_n|n\rangle. \tag{17}$$

The *transfer matrix*,

---

[2]In our calculations, we set $L_t = L/a$.

$$T \equiv e^{-a\mathcal{H}}, \qquad (18)$$

is the operator that evolves one time slice of the system into the next. Note that in this section, for simplicity, we reabsorb $\beta$ in the definition of $\mathcal{H}$. In terms of T, the partition function in Eq. (9) can be expressed as

$$Z = \mathrm{Tr}(T^{L_t}). \qquad (19)$$

Masses of particle states can be obtained from the large time decay rate of (normalized) two-point correlators of interpolating operators,

$$C(t) \equiv \frac{\langle \Omega | O^\dagger(0) O(t) | \Omega \rangle}{\langle \Omega | O^\dagger(0) O(0) | \Omega \rangle} = \frac{\langle \Omega | O^\dagger(0) T^{t/a} O(0) | \Omega \rangle}{\langle \Omega | O^\dagger(0) O(0) | \Omega \rangle}, \qquad (20)$$

where $| \Omega \rangle$ is the vacuum state, normalised so that $| \Omega \rangle = T | \Omega \rangle$, and $O(t)$ is an interpolating operator that produces the single-particle state $| \Psi \rangle$ by acting on the vacuum,

$$| \Psi \rangle = O(t) | \Omega \rangle, \qquad (21)$$

with $\langle \Omega | \Psi \rangle = 0$. Inserting a complete set of eigenstates of $\mathcal{H}$ in Eq. (20), we obtain

$$C(t) = \sum_n |c_n|^2 e^{-E_n t}, \qquad (22)$$

where the coefficients $c_n$, given by

$$c_n = \frac{\langle n | O(0) | \Omega \rangle}{\sqrt{\langle \Omega | O^\dagger(0) O(0) | \Omega \rangle}}, \qquad (23)$$

are called *overlaps*. If $E_0 < E_1 < \cdots$, then

$$C(t) \sim |c_0|^2 e^{-E_0 t} \left( 1 + \frac{|c_1|^2}{|c_0|^2} e^{-(E_1-E_0)t} + \cdots \right) \sim |c_0|^2 e^{-E_0 t},$$
$$t \gg (E_1 - E_0)^{-1}. \qquad (24)$$

Hence

$$E_0 = -\lim_{t \to \infty} \frac{1}{a} \log \frac{C(t+a)}{C(t)}. \qquad (25)$$

This equation implies that, in principle, $E_0$ can be obtained by fitting an exponential to the large $t$ values of $C(t)$ as measured from the lattice. When $O(t)$ creates a zero-momentum state, the energies $E_i$ are identified with particle masses $m_i$. In our calculation we will restrict to this case.

Following from Eq. (25), we define the effective mass $m_{\mathrm{eff}}(t)$ as

$$a m_{\mathrm{eff}}(t) \equiv -\log \frac{C(t+a)}{C(t)}. \qquad (26)$$

If a one-particle eigenstate of the Hamiltonian were propagating, $m_{\mathrm{eff}}(t)$ would be constant with respect to $t$ with a value equal to the mass of that state. In the presence of other states contributing to the correlation function, we expect this effective mass to be an upper bound for the true asymptotic mass at any finite $t$. In numerical studies, a $t_{\mathrm{min}}$ can be identified such that, for $t \geq t_{\mathrm{min}}$, only the ground state (or, more precisely, the smallest mass eigenstate with nonzero overlap) contributes to $C(t)$ within the statistical precision, and hence $a m_{\mathrm{eff}}(t)$ becomes constant. The plateau value of $m_{\mathrm{eff}}(t)$ provides an estimate of the ground state mass $m_0$, which can be extracted by fitting a single exponential to the data for $C(t)$ for $t \geq t_{\mathrm{min}}$.

While this program is at the basis of standard techniques for extracting masses from correlators, its direct implementation is not straightforward and requires a careful treatment of numerical data. The first difficulty one encounters stems from the statistical noise affecting the measurements. In fact, while the statistical fluctuations of $C(t)$ are roughly independent of $t$, the magnitude of correlation functions decays exponentially. This gives an exponentially suppressed signal-to-noise ratio which is prohibitively hard to improve upon with an increase in the measurement sample size alone. In addition, the value $t_{\mathrm{min}}$ of the onset of the single-exponential asymptotic regime is not known *a priori*; it is a model-dependent feature, sensitive to the mass spectrum in the given channel and to the choice of the operator $O$, as well as to the precision of the numerical calculation. The time $t_{\mathrm{min}}$ is extracted from the simulations. Moreover, simple arguments based on asymptotic freedom show that for a given operator and in a given channel, $t_{\mathrm{min}}$ grows exponentially as the continuum limit is approached.

The discussion above highlights the necessity to go to large times to isolate the ground state, but then the signal-to-noise ratio degrades, and this makes it difficult to estimate $m_0$ in a reliable way. If one could find operators with correlators that provide single exponential behaviors, one could perform fits at small times, when the signal is still well visible above the noise. Although this ideal situation cannot be reproduced in numerical investigations, since the knowledge of operators giving rise to single exponential correlators would only arise from an at least partial solution of the theory, one can try to engineer the calculation in such a way that in each relevant channel $t_{\mathrm{min}}$ is as small as possible. For this purpose, at each value of $a$ we construct interpolating operators that maximize the overlaps with the spectral states of interest. The main idea is to approximate the (unknown) exact eigenfunctions of $\mathcal{H}$ with an appropriate linear combination of a set of states $\{|\Psi_i\rangle\}$, chosen on the basis of symmetry considerations, as trial wave functions. Then, in the given channel, the mass of the lowest lying state above the vacuum can be bound as

$$am_0 \leq -\frac{1}{\tau} \log \left\{ \min_{\{|\Psi\rangle\}} \frac{\langle \Psi | T^\tau | \Psi \rangle}{\langle \Psi | \Psi \rangle} \right\} = am_{\text{var}}, \quad (27)$$

where $\Psi$ denotes any linear combination of the variational basis $\Psi_i$, subject to the constraint $\langle \Psi | \Omega \rangle = 0$, and $\tau$ is a time chosen for minimization, which is performed across the linear combinations of our basis operators. This bound is saturated by the lowest-lying eigenstate of the Hamiltonian in the chosen channel, which can be obtained using a complete set of variational states $\{|\Psi_i\rangle\}$. Since variational bases used in calculations are necessarily finite, the bound is in general not saturated when the variational method is used in practice. Nevertheless, with a suitably large variational basis, the extracted variational mass $m_{\text{var}}$ will eventually be compatible within the statistical errors with $m_0$. This framework, referred to henceforth as the *variational technique*, can be implemented algorithmically in order to extract both the glueball and the torelon spectrum in various channels [124].

The success of this approach and the quality of the results obtained with this technique crucially depend on the nature of the operators that we include in the variational basis. For this reason, particular attention needs to be paid to its construction. We will review in the following two subsections the approach we followed to construct trial states to be used in the variational calculation, by discussing how gauge invariant states are created on the lattice in Sec. III B, and how one obtains the irreducible representations of the symmetry group of the lattice in which these states must transform in Sec. III C. In Sec. III D we will show how to perform the extremization provided in Eq. (27) in an effective way, in order to obtain robust estimates of $m_0$. The effective description of torelon states as closed fluxtubes will be summarized in Sec. III E. The estimates of $m_0$ will be affected by systematic errors of different origins, which will be discussed in Sec. III F.

### B. States on the lattice

In this section we explain how to create gauge invariant states out of the vacuum and their interpretation in terms of glueball and torelon states. Consider traced path ordered products of links, defined in Eq. (4), around closed spacelike loops $\mathcal{C}$,

$$U(\mathcal{C}) = \text{Tr} \prod_{(x,\mu) \in \mathcal{C}} U_\mu(x), \quad (28)$$

where $\mathcal{C}$ can be defined as a set of successive displacements,

$$\mathcal{C} = [f_1, f_2, \ldots, f_L], \quad (29)$$

where each $f_j$ is one of the elementary vectors of the lattice $\{\vec{e}_i\}$. The sequence $f_1, f_2, \ldots, f_L$ is defined up to cyclic permutations. The closeness of the path $\mathcal{C}$ implies that $\sum_i f_i = 0$.

A generic gauge invariant operator $O$ such that $\langle \Omega | O | \Omega \rangle = 0$ can be obtained as a sum of products of operators $O_\mathcal{C}$, each defined as

$$O_\mathcal{C} = U(\mathcal{C}) - \langle \Omega | U(\mathcal{C}) | \Omega \rangle \quad (30)$$

for specified choices of $C$.

Single trace operators create states called glueballs when $\mathcal{C}$ is contractible and torelon states when $\mathcal{C}$ wraps around a spatial direction of the spacetime hypertorus and is thus noncontractible. These two classes of states transform in different representations of the center of the group and hence do not mix. We will start our analysis from the contractible loops. Most of our arguments are applicable also to noncontractible loops, which will be analyzed more specifically in Sec. III E.

Multitrace operators are monomials involving products of at least two of the operators in Eq. (28). Operators in this class can be used to generate multiparticle states. Some of these states have the same quantum numbers as single particle states we are interested in and can thus mix with them. This mixing can result in a systematic error in the extraction of masses of single-particle states. In our calculation, we will neglect mixing of genuine glueball states with multiparticle states. The justification for neglecting multitrace operators resides in the fact that matrix elements involving them go to zero in the large-$N$ limit.

### C. Symmetries

In the lattice theory, the Poincaré symmetry of the continuum is explicitly broken to the discrete subgroup of symmetries of the hypercubic lattice and discrete translations by an integer number of lattice spacings. In particular, on an infinite lattice, for a time slice, this is the semidirect product of discretized translations $T_d$ and of the point groups of invariance of the elementary (cubic) cell of the lattice: the *Octahedral* group $O_h$ (see, e.g., Ref. [126]). The study of the representations of this symmetry group of the lattice is simplified by the fact that $T_d$ is an invariant Abelian subgroup. The one-dimensional representations of $O_h$ (related to momentum) can thus be studied separately from those of $T_d$.

In a finite box of size $L$, and with periodic boundary conditions, the momentum is quantized in every direction as $p_n = 2\pi n/L$. On a lattice its value must also lie in the Brillouin zone containing $\vec{p} = \vec{0}$. Operators at fixed momentum can be obtained as Fourier sums of their counterpart in coordinate space,

$$O_\mathcal{C}(t, \vec{p}) = \sum_{\vec{x}} e^{i\vec{p} \cdot \vec{x}} O_\mathcal{C}(\vec{x}, t). \quad (31)$$

Zero momentum combinations (to which we restrict ourselves in this study) can be simply obtained as sums over fixed time slices of operators of the type given in Eq. (30),

$$O_{\mathcal{C}}(t, \vec{p} = \vec{0}) = \sum_{\vec{x}} O_{\mathcal{C}}(\vec{x}, t). \tag{32}$$

We now briefly describe the irreducible representations of $O_h$ and their relation with the representations of the Poincaré group. The *Octahedral* group is the symmetry group of a cube. This group has 24 elements divided into 5 conjugacy classes. Accordingly, it has 5 inequivalent irreducible representations, labeled by $R = A_1, A_2, E, T_1, T_2$, of dimensions 1, 1, 2, 3, 3, respectively. The spatial parity $P$ has two eigenstates, which we label by an additional $\pm$ sign, depending on whether they remain invariant $(+)$ or are reflected $(-)$ under a parity transformation. We will label the states of the lattice theory with $\mathcal{R} = R^P$ and their mass with $m_{R^P}$. Asterisks will denote excitations of the ground state: $A_1^{+*}$ will denote the first excited state of $A_1^+$, $A_1^{+**}$ the second, etc.

The states generated from the vacuum by gauge invariant operators $U(C)$ will transform in the same representation as the paths on which they were defined according to Eq. (28). In general, single trace operators belong to reducible representations of the octahedral group. Under the action of an element $r$ of the group, the operators $O_{\mathcal{C}}$ transform in representation $U(r)$ in the following way:

$$U(r)O_{\mathcal{C}}U^{-1}(r) = O_{r\mathcal{C}}, \tag{33}$$

where the law of transformation of $\mathcal{C}$ can be inferred from its definition in Eq. (29),

$$\mathcal{C}' = r\mathcal{C} = [rf_1, rf_2, \ldots, rf_L]. \tag{34}$$

The decomposition of $U(r)$ in terms of its irreducible components can be obtained from the orthonormality property of characters, supplemented by a choice of orthonormal bases for each of the irreducible representations $R^P$ of $O_h$. For this, the projector method borrowed from Ref. [126] has been used.

In the continuum limit, we expect the Poincaré symmetry to be recovered. The relationship between the representations of the octahedral group defined above and those of the Poincaré group enables us to decompose the former in their continuous spin components. The representations of the Poincaré group are labeled by the mass $m$ and the quantum numbers $J^{PC}$, where $J$ is associated with irreducible representations of the rotation group, $P$ with spatial parity, and $C$ with charge conjugation. Owing to the pseudoreality of the representations of $Sp(2N)$, $C$ is always positive. Hence, we will drop this quantum number from now on.

If we restrict the elements of the rotation group in a representation $J$ to the discrete rotations that lie in $O_h$, we obtain the subduced representation $J{\downarrow}O$. We report in Table I the subduced representations for the lowest values of $J$, adapted from Ref. [124]. In $O_h$, these representations are reducible in terms of $A_1$, $A_2$, $E$, $T_1$, and $T_2$. Thus, degenerate states with the same spin but different polarizations of the continuum spectrum might have a different mass on the lattice. In the continuum limit, nevertheless, the restoration of continuum rotational invariance implies that these states become degenerate. For instance, the $E$ and $T_2$ representations of the octahedral group contain, respectively, two and three of the five polarizations of spin-2 particles. Hence, corresponding states extracted in the $E^{\pm}$ and $T_{2\pm}$ channels must become degenerate in the continuum limit. The degree of degeneracy of these states at finite lattice spacing will thus provide an important measure of the effect of lattice artifacts.

TABLE I. Subduced representations $R$ of the continuum rotation group and their components labeled with the spin $J$, up to $J = 4$.

| $J$ | $A_1$ | $A_2$ | $E$ | $T_1$ | $T_2$ |
|-----|-------|-------|-----|-------|-------|
| 0 | 1 | 0 | 0 | 0 | 0 |
| 1 | 0 | 0 | 0 | 1 | 0 |
| 2 | 0 | 0 | 1 | 0 | 1 |
| 3 | 0 | 1 | 0 | 1 | 1 |
| 4 | 1 | 0 | 1 | 1 | 1 |

### D. Extraction of masses

Let us now consider a specific irreducible representation $R^P$ and build a generic linear combination $\Phi$ of basis elements $O^{R^P}$ at time $t$, which we denote as

$$\Phi(t) = \sum_i v_i O_i^{R^P}(t). \tag{35}$$

The two-point correlation function is

$$\langle \Omega | \Phi^{\dagger}(0) \Phi(t) | \Omega \rangle = \sum_{ij} v_i^{\star} v_j C_{ij}(t), \tag{36}$$

where, in general,

$$C_{ij}(t) = \sum_a c_i^{a\star} c_j^a e^{-m_a t}, \tag{37}$$

with $c_i^a = \langle a | O_i^{R^P}(0) | \Omega \rangle$. As a result, Eq. (26) can be rewritten as

$$am_{\text{eff}}(t) = -\log \frac{\sum_{ij} v_i^{\star} v_j C_{ij}(t)}{\sum_{ij} v_i^{\star} v_j C_{ij}(t-a)}. \tag{38}$$

The matrix $C_{ij}(t)$ is positive definite [see Eq. (37)], and its eigenvalues are given by $\lambda_a(t) = e^{-m_a t}$. Hence, extracting the spectrum is equivalent to the diagonalization of $C_{ij}(t)$. Unfortunately, due to statistical fluctuations, eigenvectors and masses of the measured $C_{ij}(t)$ do depend on $t$. In order to resolve this dependency, we seek a solution to the generalized eigenvalue problem

$$\sum_j C_{ij}(\tau)v_j = \lambda(\tau, 0)\sum_j C_{ij}(0)v_j, \quad (39)$$

by diagonalising $[C^{-1}(0)C(\tau)]$ for some $\tau > 0$. The eigenvectors of $[C^{-1}(0)C(\tau)]$ provide us with a practical choice $\tilde{\Phi}_i$ of the optimal operators. The corresponding masses $m_i$ can be obtained from fits of the correlators of the $\tilde{\Phi}_i$ (which we refer to as $\tilde{C}_i$) at $t > t_0$, using the ansatz

$$\tilde{C}_i(t) = 2|c_i|^2 e^{-m_i L_t a/2} \cosh m_i\left(t - \frac{L_t a}{2}\right), \quad (40)$$

over ranges of $t$ for which

$$am_{\text{eff}}(t) = \text{arccosh}\left(\frac{\tilde{C}_i(t + a) - \tilde{C}_i(t - a)}{2\tilde{C}_i(t)}\right) \quad (41)$$

reaches a plateau value. We still denote as $am_{\text{eff}}$ the effective mass, although we adopt from now on a definition that differs from the one in Eq. (26). The reason for the discrepancy, which is visible only away from the large volume limit, is a consequence of adopting periodic boundary conditions in time, which allows for both forward and backward propagating states. The mass of the ground state, $m_0$, is obtained from a fit of the largest eigenvalue $\lambda_0$. The masses of higher energy states can be obtained in the same manner from the diagonal correlators of eigenvectors associated with the other eigenvalues computed in the generalized eigenvalue problem.

As discussed earlier, a crucial ingredient for an efficient variational calculation is the preparation of trial states that have an extension comparable to the target glueball state. *A priori*, we have no information about the physical size characterizing glueball states. To determine an efficient linear combination, we shall insert in our variational set operators obtained from prototypical paths of different sizes and shapes, and also operators obtained from the original basis at each of $S$ iterations of *smearing* and *blocking* operations, with the combination obtained in Ref. [2] (to which we refer the reader also for the definition of the operations of blocking and smearing and for specific details on the particular paths used to define the basis operators). In this way, we obtain a variational basis that finely scans the propagating states from length scales corresponding to the lattice spacing all the way up to the lattice extent $L$.

From the technical perspective, the only procedural change to the methodology employed for $Sp(4)$ in Ref. [7] (to which we refer for further details) lies in the projection and cooling routines that had to be adapted to the case of $Sp(2N)$. With $M$ elementary paths and $S$ smearing steps used for constructing the basis, our variational basis is formed by $S \times M$ operators in total, and $\tilde{C}_{ij}(t)$ will accordingly have $(S \times M)(S \times M + 1)/2$ elements. In our calculations, we perform the maximum number of blocking steps $N_b$ allowed by the finite size, provided by the maximization of the left-hand side (LHS) in the inequality $2^{N_b} \leq L/a$. At each blocking step we perform 2 smearing steps and 15 cooling steps to reproject on the group. In general, our variational basis contains approximately 200 elements.

### E. Effective string theory

Torelon states are generated from the vacuum by path ordered products of link variables along noncontractible paths, i.e., paths that wrap the periodic lattice along a given direction. These states have the same quantum numbers as physical states in which a wrapping closed loop of glue with fixed length is propagating in the system. We refer to this configuration as a *fluxtube*. When the fluxtube is long enough, it can be described by an effective string theory. This classical effective theory is written in terms of a single physical parameter, the string tension $\sigma$, that governs the energy of the fluctuations. In order to extract it from the data with the highest precision, we will make use of effective string theory, as briefly summarized in this section.

Effective string theory is based on approximating the fluxtube as a one-dimensional fluctuating object—a *string*—with constant energy per unit length. Classically, the mass $m$ and the length $L$ of the fluxtube are proportional,

$$m = \sigma L. \quad (42)$$

This *classical* string description becomes exact in the *long string limit* $L^2\sigma \to \infty$.

At finite length, quantum corrections become relevant. The energy of the fluxtube is obtained as a power expansion in $1/(\sigma L^2)$ around the long string limit. In general,

$$m_0 = \sigma L\left(1 + \sum_{k=1}^{\infty}\frac{d_k}{(\sigma L^2)^k}\right), \quad (43)$$

where the dimensionless coefficients $d_k$, which are in principle calculable, can be determined by matching the power series to the results of numerical measurements. Universality theorems allow even to fix some of these coefficients on the basis of symmetry arguments. The formation of the fluxtube can be described as a process of spontaneous breaking of some of the generators of the Poincaré symmetry. We omit details, for which we refer the reader to the literature [127].

The ground state mass $m$ of a torelon wrapping along one direction of extent $L$ is given, in a spacetime of dimension $D$, by

$$m_{\text{LO}}(L) = \sigma L - \frac{\pi(D-2)}{6L}, \tag{44}$$

where we included the leading order correction in an expansion in $1/\sigma L^2$, and

$$m_{\text{NLO}}(L) = \sigma L - \frac{\pi(D-2)}{6L} - \frac{1}{2}\left(\frac{\pi(D-2)}{6}\right)^2\frac{1}{\sigma L^3}, \tag{45}$$

which describes $m(L)$ up to the next-to-leading-order correction. At this order, one can show that these predictions are *universal*; i.e., the coefficients are fixed by Poincaré invariance and certain geometric dualities. The only physical parameter to consider is thus $\sigma$.

In general, for the ground state, no deviations with respect to the Nambu-Goto formula

$$m_{\text{NG}}(L) = \sigma L\sqrt{1 - \frac{(D-2)\pi}{2\sigma L^2}} \tag{46}$$

are allowed up the term $1/(\sigma^2 L^5)$. These results will allow us to compute $\sigma$ from the mass of torelons, keeping under control the effects of working at finite $L$.

### F. Sources of systematic errors

There are several sources of systematic errors that affect the computation of glueball and torelon masses. In this section, we discuss the most relevant ones for our study.

As explained in Sec. III D, the variational technique depends on our choice of basis of operators. A potential source of error is the choice to include only single trace operators in our variational set. By doing so, we are neglecting scattering states and multitorelon states that share the quantum numbers of single glueball states of interest. In the case of scattering, we deal with states with two or more glueballs. Neglecting the interactions (an approximation that holds at large $N$), these states have masses that are about twice as large as the smallest glueball mass. Thus, below this threshold, we can safely neglect the effect of scattering states. Even above that threshold, scattering states decouple at large $N$. Multitorelon states have a mass that is in general an increasing function of $L$. Therefore, at large enough values of $L$, they decay quickly in correlators and can hence be neglected as well. The effect of these states can in principle be controlled by including the corresponding operators in the variational basis and evaluating their overlaps, as done in Ref. [124].

As a consequence of the fact that we are simulating a finite lattice, all our physical estimates will be affected by finite-size effects. These effects have been reported in Ref. [128], where it is shown that they obey the relationship

$$m(L) = m\left\{1 + \frac{be^{-\frac{\sqrt{3}}{2}mL}}{mL}\right\}, \tag{47}$$

with $m(L)$ and $m$ the masses in volume $V = L^3$ and at infinite volume, respectively. $b$ is a coefficient that, *a priori*, depends on the symmetry channel. Under the assumption that these corrections are independent of the lattice spacing $a$, we will be able to compute them at one value of $a$ and use the same prediction for all others. More so, we will be able to neglect them altogether once we find that, at a certain combination of $a$ and $L$, these effects are much smaller than the statistical error.

Discretization errors come from the dependence of the masses on the lattice spacing. A trivial dependence can be inferred from dimensional analysis. The lattice combination $ma$ is dimensionless. Since all masses obtained on the lattice depend on the lattice spacing $a$ in this way, we consider ratios of dimensionful objects where the trivial dependence simplifies in the ratio. As a reference scale for the ratio, we use the square root of the string tension $\sqrt{\sigma}$. This choice is motivated by the fact that, thanks to the results discussed in Sec. III E, we can measure the string tension more accurately than any other quantity of interest. Hence, the use of the string tension reduces significantly the systematic error due to the scale setting process, which is a necessary step to provide quantities in physical units.

Beyond the overall dependence of the mass on the lattice spacing $a$, we know, by computing the naïve continuum limit of the theory described by the lattice action in Eq. (6), that the leading corrections to mass ratios start at order $a^2$. Therefore, close to the continuum limit, for a glueball state $R^P$, we approximate

$$\frac{m_{R^P}}{\sqrt{\sigma}}(a) = \frac{m_{R^P}}{\sqrt{\sigma}}(0)(1 + c_{R^P}\sigma a^2). \tag{48}$$

To conclude this overview of systematic effects, we return to mentioning topological freezing. Near the continuum limit, the Monte Carlo updates tend to get trapped in a sector at fixed topology. This topological trapping becomes more pronounced at larger $N$ [117]. Restricting the gauge theory to a sector at fixed topology generates power-law corrections in the inverse volume that delay the onset of the large volume regime [129,130]. Both large-$N$ reduction arguments [131] and the large-$N$ scaling prescription of the $\theta$ angle [132] suggest that finite volume corrections due to topological freezing are suppressed at large $N$. The decreased severity of topological freezing as $N$ increases has been verified explicitly in Ref. [133]. In Sec. IV, we will show that topological freezing affects only a small subset of our calculations. When discussing the relevant ensembles, we shall describe how topological freezing has been accounted for in those specific cases.

### IV. NUMERICAL RESULTS

In this section, we present and discuss our main numerical results. In Sec. IV A we perform calibration and validation studies of the underlying algorithm. We also select the values of the coupling $\beta$ at which to compute the masses of torelons and glueballs for the $Sp(2N)$ Yang-Mills theories with $N = 3, 4$. In Sec. IV B, we compute the ground state mass of torelons of various lengths at fixed lattice spacing. We compare the results to the predictions discussed in Sec. III E. We also evaluate finite-size effects, alongside exposing our strategy for extracting the string tension using one lattice size, in the asymptotic regime. In Sec. IV C we report the results of the continuum limit extrapolations of the glueball spectrum for $N = 1, 2, 3, 4$, while we cover in detail in Appendix C all pertinent technicalities. The continuum limit values of the masses are then used to extrapolate toward the large-$N$ limit, in Sec. IV D.

### A. Preliminary tests and calibration studies

We compute the expectation value of the plaquette $P$, which is defined as

$$\langle P \rangle \equiv \frac{1}{6L^4} \frac{1}{2N} \sum_{x, \mu > \nu} \Re \text{Tr} \mathcal{P}_{\mu\nu}(x). \tag{49}$$

We consider several values of $\beta$, and focus attention on $N = 3$ and $N = 4$. Independent ensembles are generated for each chosen value of $\beta$, with either unit (cold) or random (hot) starting configurations in the Monte Carlo update algorithm. We calculate $\langle P \rangle$ each 5 sweeps, record $10^4$ individual measurements of this quantity out of the $5 \times 10^4$ sweeps performed for each $\beta$ value. By comparing the history of $\langle P \rangle$ starting separately with unit and random configurations, we are able to identify and discard the initial transient due to thermalization. We have verified explicitly that the integrated autocorrelation times are less than 1.5 for all values of $\beta$. We finally bin and bootstrap the measurements of $\langle P \rangle$. For $Sp(2N)$ Yang-Mills theories with $N = 3, 4$, the results are shown in Fig. 1 for lattices with size $(L/a)^4 = 16^4$.

Our algorithm is based on a heat bath update of $Sp(2)$ subgroups that when combined provide a covering of the whole $Sp(2N)$ group (see Appendix A for a detailed explanation). For validation purposes, we obtained alternative, independent estimates for the average plaquette using the simpler (and slower) Metropolis-Hastings update algorithm. For both $N = 3$ and $N = 4$, and at every relevant value of $\beta$, the estimates obtained with the two different algorithms are compatible with each other, within 1 standard deviation. For $N = 3$, independent numerical results are also available through Ref. [6], and our results are compatible with theirs within 1 standard deviation, when comparisons are possible. Finally, the limits of weak ($\beta \to \infty$) and strong ($\beta \to 0$) coupling can be controlled analytically [6]. It is expected that

$$\langle P \rangle_{\text{weak}} = 1 - \frac{(N+1)}{8\beta} + O(1/\beta^2) \tag{50}$$

and

$$\langle P \rangle_{\text{strong}} = \frac{\beta}{N} + O(\beta^5) \tag{51}$$

at weak and strong coupling, respectively. In Fig. 1, we compare the leading terms of these analytical predictions to our numerical data in the relevant regime. The combination of all these tests supports the robustness of the algorithm we are employing.

The behavior of $\langle P \rangle$ as a function of $\beta$ is also used to detect the potential presence of a bulk phase transition that separates the weak and the strong coupling regimes of the theory. While the latter is dominated by strong lattice artifacts, the former is relevant to continuum physics. The pseudoinflection point visible in Fig. 1 (for both $N = 3$ and $N = 4$) is a potential signature of such a phase transition. We study the nature of this change of regime in Appendix B, where we conclude that our numerical data are compatible with a crossover, confirming the findings of Ref. [6] for $N = 3$ and extending this conclusion to $N = 4$.

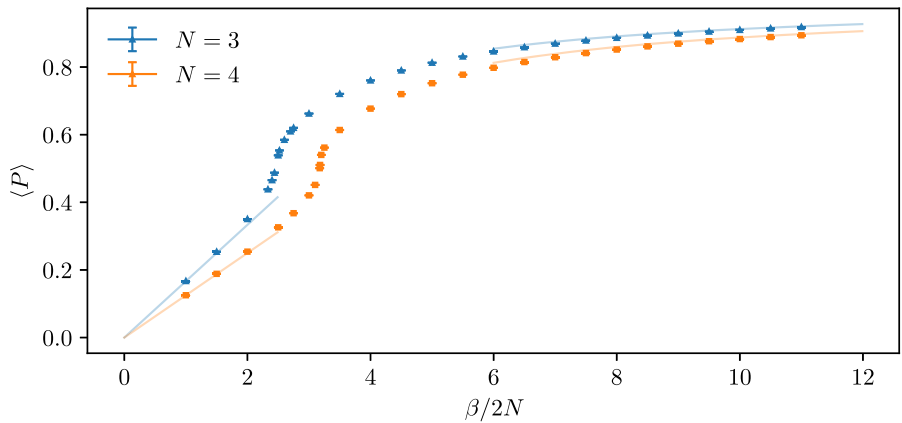

FIG. 1.   The average plaquette $\langle P \rangle$ measured at varying coupling $\beta/2N$, for fixed lattice size $L = 16a$. The results from the leading order expressions at weak ($\beta \to \infty$) and strong ($\beta \to 0$) coupling regimes in Eqs. (50) and (51) are presented by the solid lines for comparison purposes.

In principle, the absence of a genuine phase transition may allow extraction of physical observables by performing an extrapolation to the continuum limit that makes use of generic values of $\beta$. Nevertheless, by restricting our choices of $\beta$ to the weak coupling regime we maintain better control over the approach to the continuum. Our choice of the values of $\beta$ for the simulations results from a pragmatic compromise aimed at reducing discretization errors while deploying the finite amount of available computational resources.

### B. Torelons and strings

In this subsection, we discuss the methodology we use to extract the string tension from the ground state mass of torelons of length $L$ for $N = 3$, 4, while also testing the predictions of Sec. III E. We first perform an analysis of the $L$ dependence of the mass at a fixed value of the lattice coupling, in order to identify the regime in which the string effective description is applicable. We then extract the string tension from torelon masses measured at one asymptotic value of $L$ for each choice of $\beta$. This procedure allows us to estimate accurately the string tension as a function of the finite lattice size, using the known functional form of the torelon mass. We retained $10^4$ thermalized configurations for postproduction analysis. The variational basis we adopt includes the elementary Wilson line winding around a compact spatial dimension, and averaged over all three spatial directions, alongside its blocked and smeared improved versions, up to blocking level $N_b$ such that in the inequality $2^{N_b} \leq L/a$ the LHS is maximized. Following Ref. [7], to which we refer for further details, we performed either one (for the coarsest lattices) or two (for the finest lattices) smearing steps in between one blocking step.

For the study of the finite-size dependence of the torelon mass, we generated configurations with $\beta = 16.5$ for $Sp(6)$ and $\beta = 26.7$ for $Sp(8)$, on the lattice sizes listed in Table II. These values of $\beta$ are chosen to be small enough that large physical volumes are reached with moderate computing cost, while still remaining within the weak coupling regime. The values of the masses thus obtained, denoted by $am_s$, are reported in Table II and plotted in Fig. 2. In order to extract $am_s$, we performed a maximum likelihood analysis based upon Eq. (40). The value of the resulting $\chi^2/N_{\text{d.o.f.}}$ is usually below or around one; exceptions to this are mostly restricted to the largest lattice studies in $Sp(8)$, where $am_s$ becomes of order one and as a consequence the signal decays quickly.

We now test the predictions of Sec. III E. From Fig. 2, we see that, at the largest values of $L/a$, $m_s a$ is an approximately linear, increasing function of the length, in both $N = 3$ and $N = 4$. This behavior supports the intuitive description of a torelon state as a closed fluxtube with constant energy per unit length. In order to extract the string tension $\sigma$, as a first approximation we use Eq. (42) applied to the largest value of $L/a$, treating the fluxtube as a *classical* string. We call $\sigma_{\text{cl}}$ the resulting string tension. For $Sp(6)$, we find $\sigma_{\text{cl}}a^2 = 0.0212(4)$ at $L/a = 28$, and for $Sp(8)$, we obtain $\sigma_{\text{cl}}a^2 = 0.0520(14)$ at $L/a = 20$. The large-$L$ expansion is expected to be well-behaved when $\sigma L^2 \gg 1$. At a given value of $\beta$, the classical string in Eq. (42) should hence provide an accurate description of the torelon when $L \gg 7a$ for $Sp(6)$ and $L \gg 5a$ for $Sp(8)$, the numerical coefficients in these two expressions coming from the condition $\sigma L^2 \simeq 1$. Corrections to long string behavior, such as those encoded in Eqs. (44)–(46), are expected to become important as $L/a$ is decreased.

TABLE II. Ground state masses of the torelon states of $Sp(2N)$ theories for $N = 3$ and $N = 4$, at various values of $L/a$. Masses are obtained from a fit to Eq. (40).

| | $Sp(6)$, $\beta = 16.5$ | | $Sp(8)$, $\beta = 26.7$ | |
|---|---|---|---|---|
| $L/a$ | $am_s$ | $\chi^2/N_{\text{d.o.f.}}$ | $am_s$ | $\chi^2/N_{\text{d.o.f.}}$ |
| 8 | 0.1136(13) | 0.82 | 0.2108(33) | 0.54 |
| 10 | 0.1159(14) | 0.72 | 0.3963(44) | 0.33 |
| 12 | 0.1385(26) | 0.52 | 0.5644(73) | 1.0 |
| 14 | 0.2126(43) | 0.32 | 0.6609(90) | 1.03 |
| 16 | 0.3004(64) | 0.08 | 0.815(20) | 2.74 |
| 18 | 0.3372(72) | 0.71 | 0.915(11) | 1.84 |
| 20 | 0.4038(98) | 0.11 | 1.040(30) | 1.71 |
| 22 | 0.4423(90) | 1.37 | ⋯ | ⋯ |
| 24 | 0.503(11) | 0.67 | ⋯ | ⋯ |
| 28 | 0.594(11) | 1.51 | ⋯ | ⋯ |

We show in Fig. 2 our best-fit results of the numerical data, based upon Eqs. (44)–(46) and a linear form inspired by Eq. (42), restricting the fitting region to the range $L \geq 16a$ in $Sp(6)$ and $L \geq 12a$ in $Sp(8)$. The results of the fits are also reported in Table III. All the values of $\chi^2/N_{\text{d.o.f.}}$ are acceptable. Determinations based upon LO, NLO, and NG effective string treatments are indistinguishable from one another, but they are different from the classical behavior represented by the linear approximation. We elected to adopt the NG value as our best determination of the string tension as the final result of this preliminary analysis, and hence we find

$$\sigma_{N=3}a^2 = 0.02271(17), \qquad \beta = 16.5, \qquad (52)$$

for $N = 3$, and

$$\sigma_{N=4}a^2 = 0.05412(33), \qquad \beta = 26.7, \qquad (53)$$

for $N = 4$.

From this analysis, we observe that at the chosen values of $\beta$ the string picture provides a good description of the torelon mass down to lattice size $L = 16a$ for $Sp(6)$ and $L = 12a$ for $Sp(8)$. These values correspond to $L\sqrt{\sigma} \simeq 2.4$ for $Sp(6)$ and $L\sqrt{\sigma} \simeq 2.8$ for $Sp(8)$, somewhat more generous than the generic, conservative estimates we anticipated. We hence impose the bound $L\sqrt{\sigma} \geq 3$, in order to control the extraction of the string tension through an asymptotic large-$L$ expansion including at the least the LO correction. This observation will be used in the following to extract the string tension at other $\beta$ values for both $N = 3$ and $N = 4$, when we will apply the NG expression to torelon masses obtained at a single size $L$ and test *a posteriori* that $L/a$ fulfills the condition $L\sqrt{\sigma} \geq 3$.

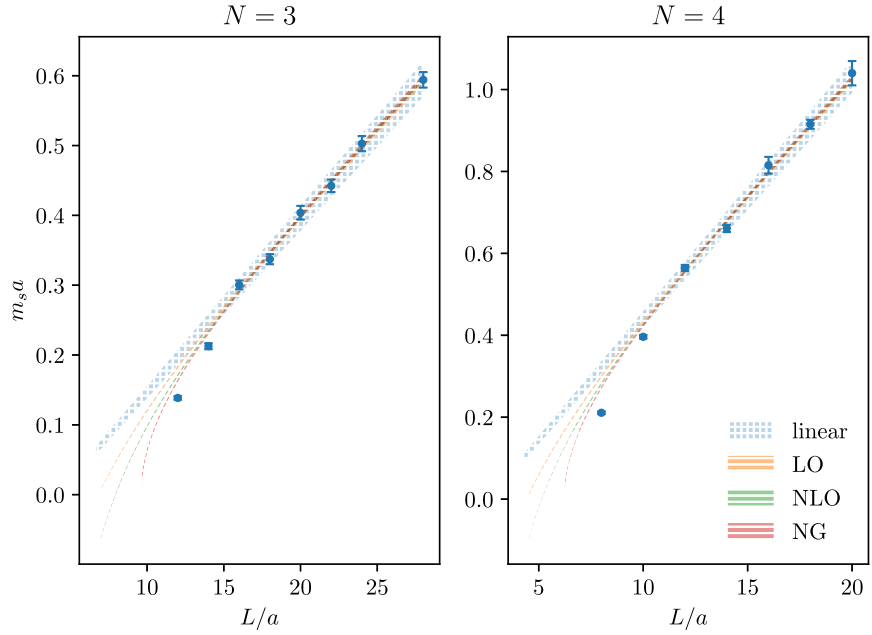

FIG. 2.   Masses of the torelons measured on lattices of volume $L^4$ at fixed lattice coupling $\beta$. We compare the results obtained by adopting a linear expression (referred to as linear, blue), leading order (yellow), next-to-leading order (green), and Nambu-Goto (red) effective description of the dependence of $m_s a$ on $L/a$. In the left-hand panel, we show the results for the $Sp(6)$ theory at coupling $\beta = 16.5$, with fits to the data in the range $L/a \geq 16$. The right-hand panel displays the results for the $Sp(8)$ theory at coupling $\beta = 26.7$, with fits to the data in the range $L/a \geq 12$. Fit curves are displayed outside the fit range in order to expose the short-$L$ deviations of the data from the asymptotic string behavior. See Table III for the fit results.

### C. The glueball spectrum

In this section, we report on the spectrum of glueballs in $Sp(2N)$ gauge theories for $N = 1, 2, 3, 4$, for each fixed value of $N$, focusing on the results obtained in the continuum. Calculations of the masses in units of the lattice spacing, finite-size effects studies, and more technical details on the continuum extrapolation can be found in Appendix C.

TABLE III.   Measurements of the string tensions $\sigma$, based upon applying Eqs. (44)–(46) and a linear form inspired by Eq. (42) to fit the dependence of numerical results of $m_s a$ on $L/a$.

|  | $Sp(6)$ $\beta = 16.5, L \geq 16a$ | | $Sp(8)$ $\beta = 26.7, L \geq 12a$ | |
|---|---|---|---|---|
|  | $\sigma a^2$ | $\chi^2/N_{\text{d.o.f.}}$ | $\sigma a^2$ | $\chi^2/N_{\text{d.o.f.}}$ |
| Linear | 0.02493(92) | 0.79 | 0.0591(20) | 1.61 |
| LO | 0.02251(17) | 0.74 | 0.05382(33) | 1.48 |
| NLO | 0.02268(17) | 0.89 | 0.05409(33) | 1.55 |
| NG | 0.02271(17) | 0.97 | 0.05412(33) | 1.59 |

We report the glueball masses in Table IV, in units of $\sqrt{\sigma}$ (top half of the table) as well as in units of the mass of the $E^+$ state (bottom half). The spectra at the various values of $N$ are also presented in Fig. 3, where, together with the lattice quantum numbers, we display the continuum quantum numbers of glueball states. The latter have been obtained from the decomposition presented in Table I under the assumption that lighter states correspond to lower continuum spin. For $N = 1$, since $Sp(2) \simeq SU(2)$, results for the spectrum are already present in the literature; see, for example, Ref. [2]. In this case, the results obtained in our study are useful for comparison and as a test of our procedure. For $SU(2)$, Ref. [2] finds the values $m_{A_1^+}/\sqrt{\sigma} = 3.78(7)$ and $m_{E^+}/\sqrt{\sigma} = 5.45(11)$. These values are compatible within 1 standard deviation with the values obtained in this work (see Table IV). For $N = 2$ we already obtained first results for the spectrum in Ref. [7]. We combine our new measurements for $Sp(4)$ with our earlier results, and in Table IV we report the weighted averages of the two. The available datasets for $Sp(4)$ are discussed in more detail in Appendix D.

A look at Fig. 3 shows that, while specific details depend on $N$, there are common patterns across the investigated values of $N$. As expected, the $A_1^+$ channel is consistently the lightest, followed by the $(T_2^+, E^+)$ (degenerate) pair. At a slightly larger mass we find the $A_1^-$ channel and the $T_2^-$ and $E^-$ (degenerate) pair. As explained in Sec. III C, the degeneracy of these pairs provides evidence that the rotation invariance of the continuum theory is recovered as $a \to 0$. The remaining channels, $A_2^\pm$ and $T_1^\pm$, are also almost degenerate in pairs, and their masses are larger than those of all other states. Since the smallest masses in the $A_2^\pm$ and $T_1^\pm$ channels are comparable with twice the ground state mass of the lowest-lying $A_1^+$ state, numerical results for these masses may be affected by systematic errors due to mixing with scattering states, as discussed in Sec. III F. An indication of this is the fact that the error bars for the masses of those heavier states are visibly larger. Large error bars are also the result of the higher level of noise affecting the extraction of masses of heavier states.

TABLE IV. Calculations of the masses in the continuum limit for each $N$ and each channel, in units of $\sqrt{\sigma}$ (top) and $m_{E^+}$ (bottom). For $N = 2$, these values have been computed as weighted means between those in Ref. [7] and those obtained in the present work; see Appendix C 2. In the case of $SU(N \to \infty)$, we have $m/\sqrt{\sigma} = 3.307(53)$ for the $A_1^{++}$ channel, 6.07(17) for the $A_1^{++}$ channel, and 4.80(14) for the $E^{++}$ channel (data taken from Ref. [2]). As expected, at least for these three channels, which are the only ones for which we can compare, the masses of $Sp(N \to \infty)$ and $SU(N \to \infty)$ theories are compatible.

| | 1 | 2 | 3 | 4 | ∞ |
|---|---|---|---|---|---|
| $R^P$ | $m_{R^P}/\sqrt{\sigma}$ | $m_{R^P}/\sqrt{\sigma}$ | $m_{R^P}/\sqrt{\sigma}$ | $m_{R^P}/\sqrt{\sigma}$ | $m_{R^P}/\sqrt{\sigma}$ |
| $A_1^+$ | 3.841(84) | 3.577(49) | 3.430(75) | 3.308(98) | 3.241(88) |
| $A_1^{+*}$ | 5.22(33) | 6.049(40) | 5.63(32) | 5.58(44) | 6.29(33) |
| $A_1^-$ | 6.20(14) | 5.69(16) | 5.22(23) | 5.36(26) | 5.00(22) |
| $A_1^{-*}$ | 7.37(72) | 7.809(79) | 6.59(49) | 7.76(85) | 7.31(45) |
| $A_2^+$ | 6.81(31) | 7.91(16) | 7.36(39) | 6.5(1.0) | 8.22(46) |
| $A_2^-$ | 8.99(86) | 9.30(38) | 8.60(67) | 7.2(1.4) | 8.69(83) |
| $T_2^+$ | 5.29(20) | 5.050(88) | 5.09(16) | 4.73(23) | 4.80(20) |
| $T_2^-$ | 6.55(34) | 6.879(88) | 6.47(43) | 6.36(35) | 6.71(35) |
| $E^+$ | 5.33(18) | 5.05(13) | 5.03(13) | 4.62(29) | 4.79(19) |
| $E^-$ | 6.61(37) | 6.65(12) | 6.34(40) | 6.29(29) | 6.44(33) |
| $T_1^+$ | 8.58(41) | 8.67(28) | 7.77(59) | 8.45(52) | 8.33(51) |
| $T_1^-$ | 9.63(77) | 9.24(33) | 9.15(69) | 8.90(75) | 8.76(72) |
| $R^P$ | $m_{R^P}/m_{E^+}$ | $m_{R^P}/m_{E^+}$ | $m_{R^P}/m_{E^+}$ | $m_{R^P}/m_{E^+}$ | $m_{R^P}/m_{E^+}$ |
| $A_1^+$ | 0.710(33) | 0.711(21) | 0.674(23) | 0.708(44) | 0.678(32) |
| $A_1^{+*}$ | 0.957(77) | 1.199(37) | 1.110(70) | 1.20(11) | 1.275(81) |
| $A_1^-$ | 1.159(54) | 1.123(44) | 1.019(57) | 1.118(87) | 1.008(66) |
| $A_1^{-*}$ | 1.40(10) | 1.541(47) | 1.41(11) | 1.57(18) | 1.55(12) |
| $A_2^+$ | 1.264(79) | 1.573(57) | 1.437(97) | 1.44(29) | 1.66(12) |
| $A_2^-$ | 1.66(18) | 1.850(94) | 1.76(14) | 1.63(38) | 1.79(18) |
| $T_2^+$ | 0.968(56) | 1.003(34) | 1.008(41) | 1.049(75) | 1.046(56) |
| $T_2^-$ | 1.223(85) | 1.375(45) | 1.307(94) | 1.44(11) | 1.454(96) |
| $E^+$ | ... | ... | ... | ... | 1.00(65) |
| $E^-$ | 1.235(99) | 1.330(44) | 1.310(95) | 1.37(12) | 1.41(10) |
| $T_1^+$ | 1.59(11) | 1.707(74) | 1.50(13) | 1.87(16) | 1.76(13) |
| $T_1^-$ | 1.85(18) | 1.820(86) | 1.80(15) | 2.01(20) | 1.84(17) |

We were able to extract masses of excited states for the $A_1^\pm$ channels at all values of $N$. These masses are reported in Table IV and displayed above the corresponding ground states in Fig. 3. The error bars of the $A_1^{+*}$ states are comparable to those of the ground state in the $A_1^-$ channel, while for the $A_1^{-*}$ states they are similar to those found in heavier channels.

Finally, we note that, where determined in both calculations, corresponding states obtained from a recent $SU(3)$ study [134] are in broad agreement with the spectrum resulting from our investigation.

### D. The spectrum toward the large-$N$ limit

As shown, for instance, in Ref. [135], while corresponding quantities in $SU(N)$ and $Sp(2N)$ Yang-Mills theories converge to a common large-$N$ limit, the $1/N$ expansions around this limit are different: in the case of $SU(N)$, only even powers of $1/N$ are present, while for $Sp(2N)$ the power expansion is genuinely in $1/N$. Following the strategy that has been implemented in the large-$N$ extrapolation of the $SU(N)$ glueball masses, we shall investigate whether the lowest order correction to the large-$N$ limit is sufficient to describe the large-$N$ glueball spectrum in $Sp(2N)$ for all the simulated values of $N$. Therefore, we fit the finite-$N$ spectrum with the ansatz

$$\frac{m_{R^P}}{\sqrt{\sigma}}(N) = \frac{m_{R^P}}{\sqrt{\sigma}}(\infty) + \frac{c_{R^P}}{N}, \tag{54}$$

where $c_{R^P}$ is a constant (expected to be of order 1 in a well-behaved expansion) that depends on the glueball channel. If the ansatz provides a sufficiently accurate description of the data, $\frac{m_{R^P}}{\sqrt{\sigma}}(\infty)$ is a reliable infinite-$N$ extrapolation of the ratio of the mass in the channel $R^P$ normalized to the square root of the string tension.

For each channel, we perform a separate linear fit to Eq. (54) using $c_{R^P}$ and $m_{R^P}/\sqrt{\sigma}(\infty)$ as fitting parameters. The results of the fits are reported in Table V. The fitting range includes all the values of $N$. From Fig. 4 we see that Eq. (54) describes the data well in this range of $N$ for the $A_1^\pm$ channels, for the $T_2^\pm$, $E^\pm$ degenerate pairs, and for the $T_1^\pm$ channels. For the $A_1^{-*}$ and for the $A_2^\pm$ channels, the value of $\chi^2/N_{\text{d.o.f.}}$ is larger than the critical value at the 5% confidence level. For comparison, in Appendix E, the same fits are performed for a range $N > 1$. Although generally the $\chi^2/N_{\text{d.o.f.}}$ are smaller, in this latter case the parameters $c_{R^P}$ and $m_{R^P}/\sqrt{\sigma}(\infty)$ are estimated from three data points only and thus only 1 degree of freedom remains to assess the goodness of the fits. For this reason, we opt to present the extrapolations including the (generally still acceptable) $N = 1$ data points, postponing to future studies that investigate larger $N$ the question of whether $N = 1$ is captured by a simple leading correction with the current precision of the data. For the time being, in the absence of any evidence that would suggest otherwise, we assume that indeed this is the case.

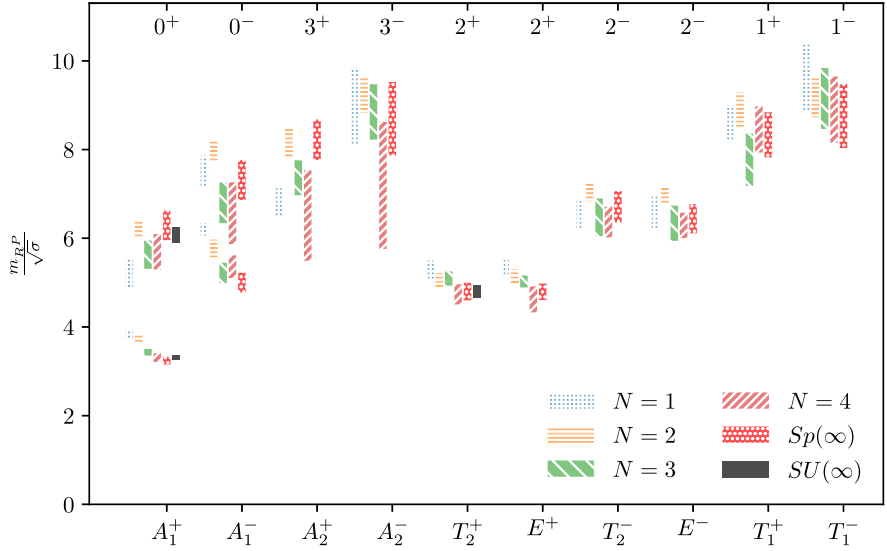

FIG. 3. Spectrum of the $Sp(2N)$ theory in the continuum limit for $N = 1, 2, 3, 4$, and $N = \infty$, in units of $\sqrt{\sigma}$. Continuum quantum numbers are reported at the top. For comparison, we have reported also the masses of the $A_1^{++}$, $A_1^{++*}$, and $E^{++}$ states for $SU(\infty)$ (borrowed from Ref. [2]). The boxes represent $1\sigma$ statistical errors.

For some of the lightest states for which the continuum mass in the large-$N$ limit is available in the literature, we can verify that the large-$N$ extrapolation of the $Sp(2N)$ and of the $SU(N)$ values are compatible. In Fig. 3 the spectrum in the large-$N$ limit is represented together with the finite-$N$ one, to allow for such a comparison. Recalling that charge conjugation is always positive in $Sp(2N)$, for the sake of comparing corresponding states, we temporarily reintroduce the corresponding index in the notation for glueball states for the rest of the current subsection. With the second $+$ superscript identifying positive charge conjugation, we borrow the values of the $A_1^{++}$, $A_1^{++*}$, and $E^{++}$ channel masses for $SU(\infty)$ from Ref. [2]. Figure 3 shows that the large-$N$ extrapolations of the $A_1^{++}$, the $A_1^{++*}$, and the $E^{++}$ in $Sp(2N)$ and $SU(N)$ are compatible with each other, as predicted by general large-$N$ arguments.

Armed with the results of the mass calculation of the $A_1^{++}$, we can provide further support to the conjecture put forward in Ref. [86] that the quantity $\eta$ in the relationship

$$\frac{m_{0^{++}}^2}{\sigma} = \eta \frac{C_2(A)}{C_2(F)} \tag{55}$$

is a universal constant depending only on the dimension of spacetime. In this equation, $C_2(F)$ and $C_2(A)$ are the quadratic Casimir operators in the fundamental and adjoint representations, respectively, whose ratio in $Sp(2N)$ is given by

$$\frac{C_2(A)}{C_2(F)} = \frac{4(N+1)}{2N+1}. \tag{56}$$

After performing the standard identification of the $A_1^{++}$ with the lowest-lying scalar glueball, we tested this conjecture by performing a fit of Eq. (55) to the data, using $\eta$ as a fitting parameter. The result can be found in Table VI and is represented in the top panel of Fig. 5. The behavior of $\eta$ as a function of $N$ is compatible with a constant for both the $Sp(2N)$ and the $SU(N)$ sequences. Moreover, the values of $\eta$ extracted in each sequence are compatible with each other within 1 standard deviation, as reported in Table VI. As an additional test of Eq. (55), the behavior of $m_{0^{++}}^2/(\sqrt{\sigma}\eta)$ is represented in Fig. 5 for both $Sp(2N)$ and $SU(N)$ groups, along with the ratio of the quadratic Casimir operators. The weighted mean of the values of $m_{0^{++}}^2/(\sqrt{\sigma}\eta)$ obtained in each series is also reported in Table VI and represented in Fig. 5. This analysis provides further indications of the validity of the conjectured Casimir scaling, at least within current accuracy and precision.

TABLE V. Large-$N$ extrapolated masses of the glueball spectrum obtained from a fit of Eq. (54). Note that the left-hand part of this table is the same as the last column of Table IV.

| $R^P$ | $m_{R^P}/\sqrt{\sigma}(N = \infty)$ | $c_{R^P}$ | $\chi^2/N_{\text{d.o.f.}}$ |
|---|---|---|---|
| $A_1^+$ | 3.241(88) | 1.29(29) | 2.38 |
| $A_1^{+*}$ | 6.29(33) | −1.6(1.2) | 2.91 |
| $A_1^-$ | 5.00(22) | 2.43(60) | 0.63 |
| $A_1^{-*}$ | 7.31(45) | 0.9(1.4) | 3.5 |
| $A_2^+$ | 8.22(46) | −2.5(1.3) | 3.3 |
| $A_2^-$ | 8.69(83) | 1.3(3.0) | 0.9 |
| $T_2^+$ | 4.80(20) | 1.01(69) | 0.65 |
| $E^+$ | 4.79(19) | 1.15(63) | 0.72 |
| $T_2^-$ | 6.71(35) | 0.1(1.2) | 1.97 |
| $E^-$ | 6.44(33) | 0.9(1.2) | 2.03 |
| $T_1^+$ | 8.33(51) | 0.7(1.6) | 1.15 |
| $T_1^-$ | 8.76(72) | 1.7(2.6) | 0.02 |

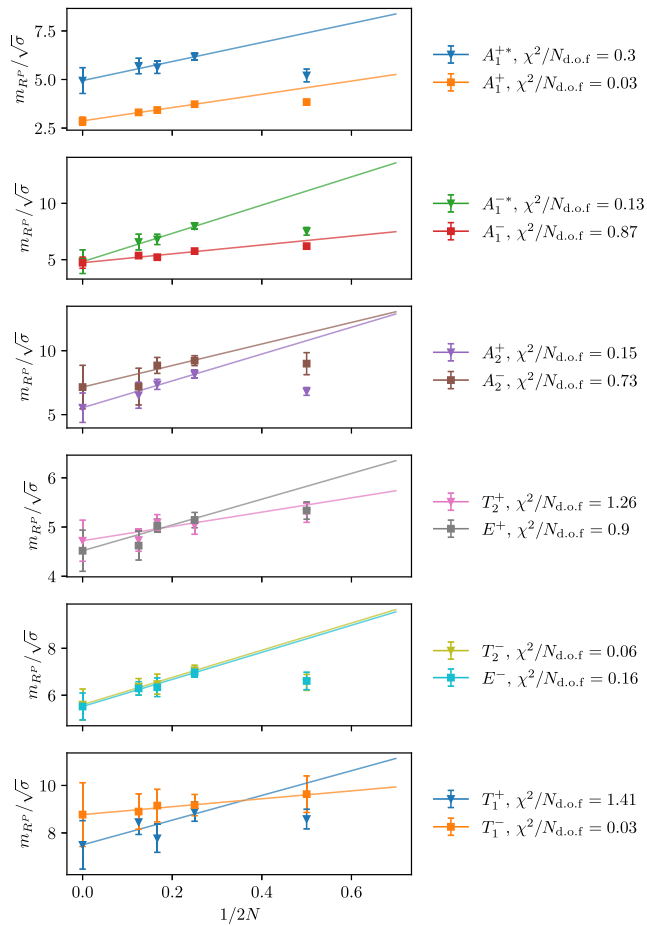

FIG. 4. Glueball mass in each symmetry channel $R^P$ in units of $\sqrt{\sigma}$, as a function of $1/2N$. The point corresponding to $1/2N = 0$ is the value of $m_{R^P}/\sqrt{\sigma}(\infty)$ obtained from the best fit of Eq. (54) to the numerical measurements reported in this publication. See main text for details.

Another remarkable universal property is the independence on the gauge group of the ratio between the mass of the tensor glueball and the mass of the scalar glueball. This has been the subject of the investigation reported in Ref. [83] that makes use of the measurements reported here. We do not repeat the details of that analysis, but refer the interested reader to Ref. [83].

TABLE VI. Resulting values of the universal constant $\eta$ for the Casimir scaling described in Eq. (55) for $Sp(2N)$ and $SU(N)$ groups. The combined fit to both groups is also reported.

| Group | $\eta(0^+)$ | $\chi^2/N_{\text{d.o.f.}}$ |
|---|---|---|
| $SU(N)$ | 5.41(10) | 1.43 |
| $Sp(2N)$ | 5.35(13) | 2.02 |
| Combined | 5.388(81) | 1.49 |

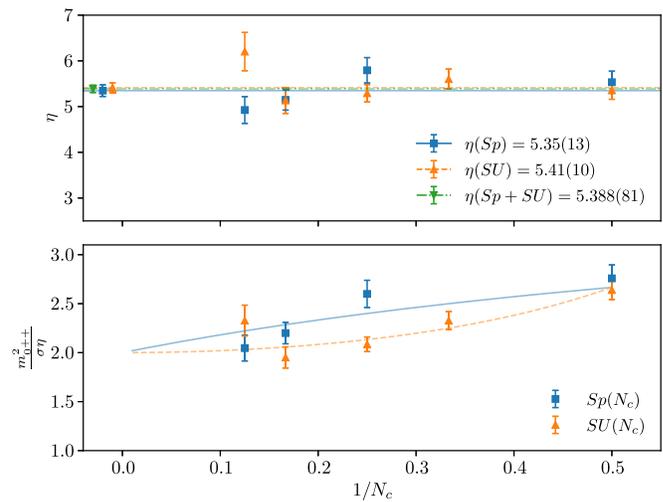

FIG. 5. Top panel: ratios defining the conjectured universal constant $\eta$ for both $SU(N_c = N)$ and $Sp(N_c = 2N)$. Note that the naming convention for the symplectic group has been altered, using the variable $N_c = 2N$, to better accommodate the data into the plots; fits are also shown for the $Sp(N_c)$ family, the $SU(N_c)$ family, and the combination of $Sp(N_c)$ and $SU(N_c)$ results. Bottom panel: measured ratios $m_{0^{++}}^2/\sigma$ further divided by the fitted universality constant $\eta$ plotted as a function of $1/N_c$; lines are the ratios of the quadratic Casimir operators of the adjoint representation over the corresponding ones of the fundamental representation as $N_c$ varies. (We note that, for the sake of the visualization, in this figure we have represented $N_c$ as a real number.)

## V. CONCLUSIONS

We have performed a numerical study of the low-lying spectrum in $Sp(2N)$ Yang-Mills gauge theories. We have considered the lattice theory formulated with $N = 1, 2, 3, 4$, and we have measured numerically its torelon and low-lying glueball spectrum as a function of the lattice coupling $\beta$. After estimating finite-size effects on the target observables by using effective-string-theory motivated predictions applied to torelon masses at $N = 3, 4$, we have extracted the string tension as a function of $\beta$ and $N$ from the latter quantities. As a by-product, through this calculation we have explicitly verified the realization of the confinement scenario in $Sp(6)$ and $Sp(8)$ Yang-Mills theories by exposing one of its most typical signatures: the presence of stringy states wrapping compact directions. While this is hardly surprising, direct validation of the expected behavior in these two gauge theories had never been done before in the literature. We have then extrapolated to the continuum limit the measurements of the adimensional ratios between the glueball masses and the square root of the string tension. Finally, we have obtained the large-$N$ limit of the glueball spectrum in the $Sp(2N)$ sequence of groups through an extrapolation in a power series in $1/N$. For the lowest-lying masses, the leading corrections $\mathcal{O}(1/N)$ to the large-$N$ limit appear to be sufficient to describe the $N$ dependence down to the smallest value $N = 1$. We have assessed the size of systematic errors connected with the continuum and the large-$N$ extrapolations and showed that they are negligible at the level of precision of our data.

We have found that, for the states for which the large-$N$ extrapolation in the $SU(N)$ sequence has been measured, their masses in the large-$N$ limit agree with the ones we have extracted taking the same limit in the $Sp(2N)$ sequence, as expected. The other states we have determined in this calculation extend our knowledge of the continuum large-$N$ spectrum, therefore providing a more complete set of masses to compare to analytic methods that naturally work at $N = \infty$, such as gauge-gravity duality techniques. Through an analysis of the ratio of the lowest-lying glueball mass squared and the string tension as a function of $N$, we have provided further support to the conjecture put forward in Ref. [86] that this ratio is proportional to the ratio of the quadratic Casimir of the adjoint over that of the fundamental representation of the gauge group. Indeed, we have verified that the $m_{0^{++}}^2/\sigma$ ratio normalized with the appropriate ratio of quadratic Casimir operators is constant within the $Sp(2N)$ and the $SU(N)$ family, and takes compatible values in the two. Our calculation bounds possible $N$-dependent corrections to this constant to be less than 10%, the latter being the minimum precision with which we have measured the ratio as a function of $N$.

In addition to the glueball spectrum at finite $N \leq 4$, our study has also provided a preliminary investigation of the topological charge in $Sp(2N)$ gauge theories, in relation to systematics effects in the generation of configurations and in the extraction of spectral masses. An extended analysis of topological observables and a more thorough analysis of topological freezing effects at large $N$ is currently in progress and will be reported elsewhere.

We envision a number of possible future avenues for exploration, in order to improve and extend this study. Beside the obvious increase in precision that can be obtained by simulating at larger $N$ and smaller lattice spacing (both of which, however, are affected by increased autocorrelation times near the continuum limit and as $N$ grows), one could investigate the effect of including double-torelon and scattering states in the operator basis, in order to have a better resolution of genuine glueball masses. A study of glueball scattering would also provide an extension to the physical reach of our current investigation (see, for example, Refs. [136,137], for the case of scalar glueball scattering in $SU(2)$ lattice gauge theory). Indeed, a scenario in which $Sp(2N)$ glueballs may play a central role is gluonic dark matter [138,139]. In order to assess the viability of a dark matter scenario based on $Sp(2N)$ glueballs, one would have to compute the cross section for the decay of the higher-spin glueballs into two scalar glueballs. This (very challenging) calculation would require a dedicated study of multipoint glueball functions. A study of correlators describing glueball scattering would be a natural starting point for such an investigation. As observed in the context of QCD (see, e.g., Ref. [140]), we expect that the presence of fundamental dynamical fermions does not shift significantly the masses in the glueball spectrum. Moreover, the mild $N$ dependence in the gluonic observables provides a first indication that no large variations will emerge across corresponding relevant physical observables evaluated in different $Sp(2N)$ gauge theories, as long as the theory is dominated by gluon dynamics, with small numbers of matter fields.

Finally, it is worth reminding the reader that the main motivation for our work has been provided by our ongoing investigation of the pseudo-Nambu-Goldstone-boson mechanism of electroweak symmetry breaking based on the $SU(4) \mapsto Sp(4)$ global symmetry breaking pattern in $Sp(2N)$ gauge theories with two fundamental Dirac flavors, following the program outlined in Ref. [7]. In this context, it was shown in Ref. [81] that the meson spectrum in the theory with dynamical fundamental fermions is well approximated by the quenched results. In principle, the calculation of the meson spectrum of $Sp(2N)$ gauge theories allows one to probe the extent and the bounds of validity of this similarity. The present work may be considered to contribute to this line of research by providing a reference energy scale for a comparison with the fermionic matter spectrum, both in the quenched and in the full theories.

## ACKNOWLEDGMENTS

The work of E. B. has been funded by the Supercomputing Wales project, which is part-funded by the European Regional Development Fund (ERDF) via Welsh Government. J. H. is supported by the STFC Consolidated Grant No. ST/P00055X/1, by the College of Science, Swansea University, and by the Grant No. STFC-DTG ST/R505158/1. The work of D. K. H. was supported by Basic Science Research Program through the National Research Foundation of Korea (NRF) funded by the Ministry of Education (NRF-2017R1D1A1B06033701). The work of J. W. L. is supported in part by the National Research Foundation of Korea funded by the Ministry of Science and ICT (NRF-2018R1C1B3001379) and in part by Korea Research Fellowship program funded by the Ministry of Science, ICT and Future Planning through the National Research Foundation of Korea (2016H1D3A1909283). The work of C. J. D. L. is supported by the Taiwanese MoST Grant No. 105-2628-M-009-003-MY4. The work of B. L. and M. P. has been supported in part by the STFC Consolidated Grants No. ST/P00055X/1 and No. ST/T000813/1. B. L. and M. P. received funding from the European Research Council (ERC) under the European Union's Horizon 2020 research and innovation program under Grant Agreement No. 813942. The work of B. L. is further supported in part by the Royal Society Wolfson Research Merit Award No. WM170010 and by the Leverhulme Trust Research Fellowship No. RF-2020-461 \9. The work of D. V. is supported in part by the INFN HPC-HTC project and in part by the Simons Foundation under the program "Targeted Grants to Institutes" awarded to the Hamilton Mathematics Institute. D. V. thanks C. Bonati, M. D'Elia, and L. Gallina for useful discussions. Numerical simulations have been performed on the Swansea SUNBIRD system, on the local HPC clusters in Pusan National University (PNU) and in National Chiao-Tung University (NCTU), and on the Cambridge Service for Data Driven Discovery (CSD3). The Swansea SUNBIRD system is part of the Supercomputing Wales project, which is part funded by the European Regional Development Fund (ERDF) via Welsh Government. CSD3 is operated in part by the University of Cambridge Research Computing on behalf of the STFC DiRAC HPC Facility (www.dirac.ac.uk). The DiRAC component of CSD3 was funded by BEIS capital funding via STFC capital Grants No. ST/P002307/1 and No. ST/R002452/1 and STFC operations Grant No. ST/R00689X/1. DiRAC is part of the National e-Infrastructure.

## APPENDIX A: CABIBBO-MARINARI UPDATING FOR $Sp(2N)$

The $Sp(2N)$ group is the subgroup of $SU(2N)$ with elements $U$ satisfying the relationship

$$U\Omega U^T = \Omega, \tag{A1}$$

where the superscript $T$ indicates the transposition operation and $\Omega$ is the symplectic matrix. The latter can be cast in the form

$$\Omega = \begin{pmatrix} 0 & \mathbb{1} \\ -\mathbb{1} & 0 \end{pmatrix}, \tag{A2}$$

with 1 the $N \times N$ identity matrix. Equation (A1) implies that $U$ has the form

$$U = \begin{pmatrix} A & B \\ -B^* & A^* \end{pmatrix}, \tag{A3}$$

with the $N \times N$ matrices $A$ and $B$ satisfying the conditions $A^\dagger A + B^\dagger B = \mathbb{1}$ and $A^T B = B^T A$.

As briefly mentioned in Sec. II, ensembles of $Sp(2N)$ configurations distributed according to Eq. (6) are obtained from lattice sweeps of single link HB and OR updates. In our implementation of these algorithms, we have used an adaptation of the Cabibbo-Marinari method [109] to the case of $Sp(2N)$. The Cabibbo-Marinari algorithm updates a group matrix via subsequent updates of $SU(2)$ subgroups covering the whole target gauge group. The choice of the set of $SU(2)$ subgroups to update is crucial. For $SU(N)$, an efficient implementation can be obtained starting from all the Cartan generators $(i, j)$ having 1 on the $i$th diagonal element, $-1$ on the $j$th diagonal element (with $1 \le i < j \le N$), and 0 everywhere else, along with their eigenvectors under conjugate action. Each generates an $SU(2)$ subgroup of $SU(N)$. Since $Sp(2N)$ is a subgroup of $SU(2N)$, the desired set of subgroups can be obtained from the set found for $SU(2N)$ by excluding the $SU(2)$ subgroups that alter the block structure in Eq. (A3) of the $Sp(2N)$ matrices. Chosing a larger set improves the decorrelation of the algorithm. In this work, we used $N^2$ subgroups.

To better understand how these subgroups are embedded in $Sp(2N)$ matrices, we reformulate the considerations above in the language of group representations. Each choice of Cartan generators, along with its eigenvectors, exponentiates to a $SU(2)$ subgroup of $SU(N)$. The elements of the matrices of this subgroup are different from a unit matrix only at the positions $\{(i, j), (j, i), (i, i), (j, j)\}$. A $SU(2)$ matrix is thus *embedded* into a $SU(N)$ matrix. We denote this embedding with $(i, j)$. The different embeddings $(i, j)$ can be seen as completely reducible representations of $SU(2)$ that are unitarily equivalent to $R \oplus \mathbb{1}_{N-2,N-2}$, i.e., to the $(1,2)$ embedding, where $R$ is the $2 \times 2$ irreducible representation of $SU(2)$. The unitary transformation that maps one embedding into another is the exchange of rows and columns $i$ and $j$ with 1 and 2, respectively. If $[N]_{\mathrm{SU}}$ is the fundamental representation of $SU(N)$ and $\{2\}$ the fundamental representation of $SU(2)$, then all the embeddings above can be decomposed as

$$[N]_{SU} = \{2\} \oplus (N - 2)\mathbb{1}. \tag{A4}$$

For the $Sp(2N)$ case, the allowed $SU(2)$ embeddings must respect the block structure Eq. (A3). These embeddings can be split into three classes that are not unitarily equivalent.

The embedding $(1,2)$ is unitarily equivalent to the embeddings $(i < N, j < N)$. Embeddings in this class can be denoted by

$$[2N]_{Sp} = \{2\} \oplus \{2\} \oplus (2N - 4)\mathbb{1}. \tag{A5}$$

Examples are, for $N = 3$,

$$
\begin{pmatrix}
1 & 0 & 0 & 0 & 0 & 0 \\
0 & a & b & 0 & 0 & 0 \\
0 & -b^* & a^* & 0 & 0 & 0 \\
0 & 0 & 0 & 1 & 0 & 0 \\
0 & 0 & 0 & 0 & a^* & b^* \\
0 & 0 & 0 & 0 & -b & a
\end{pmatrix},
\begin{pmatrix}
a & 0 & b & 0 & 0 & 0 \\
0 & 1 & 0 & 0 & 0 & 0 \\
-b^* & 0 & a^* & 0 & 0 & 0 \\
0 & 0 & 0 & a^* & 0 & b^* \\
0 & 0 & 0 & 0 & 1 & 0 \\
0 & 0 & 0 & -b & 0 & a
\end{pmatrix}, \ldots,
\tag{A6}
$$

with $a, b \in \mathbb{C}$ such that $|a|^2 + |b|^2 = 1$ and $a^*b - b^*a = 0$. There are $N(N-1)/2$ of those embeddings.

The embedding $(1, 2)'$ is unitarily equivalent to the embeddings $(i < N, j < N)'$. Embeddings in this class can be denoted by

$$
[2N]'_{Sp} = \{2\}' \oplus \{2\}' \oplus (2N-4)\mathbb{1}.
\tag{A7}
$$

Examples are, for $N = 3$,

$$
\begin{pmatrix}
1 & 0 & 0 & 0 & 0 & 0 \\
0 & a & 0 & 0 & 0 & b \\
0 & 0 & -a^* & 0 & -b^* & 0 \\
0 & 0 & 0 & 1 & 0 & 0 \\
0 & 0 & -b^* & 0 & a^* & 0 \\
0 & b & 0 & 0 & 0 & -a
\end{pmatrix},
\begin{pmatrix}
a & 0 & 0 & 0 & 0 & b \\
0 & 1 & 0 & 0 & 0 & 0 \\
0 & 0 & -a^* & -b^* & 0 & 0 \\
0 & 0 & -b^* & a^* & 0 & 0 \\
0 & 0 & 0 & 0 & 1 & 0 \\
b & 0 & 0 & 0 & 0 & -a
\end{pmatrix}, \ldots,
\tag{A8}
$$

with $a, b \in \mathbb{C}$ such that $|a|^2 + |b|^2 = 1$ and $a^*b - b^*a = 0$. There are $N(N-1)/2$ of those embeddings.

The embedding $(1, 1 + N)$ is unitarily equivalent to the embeddings $(i, i + N)$. These can be denoted by

$$
[2N]_{Sp} = \{2\} \oplus (2N-2)\mathbb{1}.
\tag{A9}
$$

Examples are, for $N = 3$,

$$
\begin{pmatrix}
1 & 0 & 0 & 0 & 0 & 0 \\
0 & a & 0 & 0 & b & 0 \\
0 & 0 & 1 & 0 & 0 & 0 \\
0 & 0 & 0 & 1 & 0 & 0 \\
0 & -b^* & 0 & 0 & a^* & 0 \\
0 & 0 & 0 & 0 & 0 & 1
\end{pmatrix},
\begin{pmatrix}
a & 0 & 0 & b & 0 & 0 \\
0 & 1 & 0 & 0 & 0 & 0 \\
0 & 0 & 1 & 0 & 0 & 0 \\
-b^* & 0 & 0 & a^* & 0 & 0 \\
0 & 0 & 0 & 0 & 1 & 0 \\
0 & 0 & 0 & 0 & 0 & 1
\end{pmatrix}, \ldots,
\tag{A10}
$$

with $a, b \in \mathbb{C}$ such that $|a|^2 + |b|^2 = 1$ and $a^*b - b^*a = 0$. There are $N$ of those embeddings.

With our construction, we have identified $N^2$ embeddings that cover the whole of $Sp(2N)$. A HB iteration on one link consists in updating consecutively each of the embeddings belonging to classes Eqs. (A5), (A7), and (A9) with the Creutz or Kennedy-Pendleton implementation of the $SU(2)$ HB algorithm. An OR iteration is built in a similar way.

For this work, we performed one HB iteration followed by 4 OR iterations for each link variable. Repeating these iterations for all the links of the lattice is a *lattice sweep*. To prevent the desymplectization and deunitarization of the configuration caused by the accumulation of numerical error, we reprojected each link of the configuration on the group after each ten lattice sweeps with a modified Gram-Schmidt algorithm that preserves the $Sp(2N)$ structure [7].

## APPENDIX B: SEARCHING FOR THE BULK PHASE TRANSITION

A phase transition taking place in the $(d + 1)$-dimensional classical canonical system defined by Eq. (9) is called a *bulk* phase transition. This transition separates the strong and weak coupling regimes of the theory, limiting the range of $\beta$ that is analytically connected to the continuum limit. The identification of bulk phase transition points is hence a crucial step for extrapolating numerical data to the $a \to 0$ limit in a controlled way.

In general terms, a bulk phase transition takes place at values of $\beta$ for which one of the derivatives of $Z(\beta)$ with respect to $\beta$ is singular (in the $L \to \infty$ limit). For a system defined by Eq. (9), with the action in Eq. (6), the first derivative of $\ln Z(\beta)$ corresponds to the expectation value of the average plaquette,

$$\langle P(\beta) \rangle \equiv \frac{1}{L^4} \frac{\partial \ln Z(\beta)}{\partial \beta}, \qquad (B1)$$

and its second derivative to the plaquette susceptibility,

$$\chi_p(\beta) \equiv \frac{\partial^2 \ln Z(\beta)}{\partial \beta^2} = L^4 [\langle P^2(\beta) \rangle - \langle P(\beta) \rangle^2]. \qquad (B2)$$

As we observed in Sec. IV A, $\langle P \rangle (\beta)$ shows a pseudoinflection point at some value $\beta_c$ of the lattice coupling. This pseudoinflection corresponds to a maximum $\chi_c(L)$ of the plaquette susceptibility. If the latter is associated with the smoothing of a proper phase transition, we expect $\chi_c(L) \to \infty$, and $\beta_c(L) \to \beta_c$, as $L \to \infty$. Conversely, if $\chi_c(L)$ stays finite when $L \to \infty$, the change from the strong coupling regime to the weak coupling one happens not through a phase transition, but via a crossover.[3]

To study the scaling of the maximum of the plaquette susceptibility with the size of the system, we focused our attention on the neighborhood of the identified pseudocritical coupling $\beta_c$, computing at various values of $\beta$ near this coupling for $L/a = 8$, 12, 16, and collecting measurements of $\langle P \rangle (\beta)$. A total of 3000 data points at each investigated value of $\beta$ and $L$ were collected, one every five sweeps. The corresponding results for $\langle P \rangle (\beta)$ are reported in Table VII. The Monte Carlo histories of our simulations were searched for any sign of metastability, which would have signaled a first order phase transition, with negative results. This allows us to exclude the presence of a discontinuous phase transition for both $N = 3$ and

TABLE VII. Expectation values of the plaquette $\langle P(\beta) \rangle$ in $Sp(2N)$ lattice theories with $N = 3$ (left) and $N = 4$ (right) at $L = 16a$.

| $\beta/2N$ | $\langle P(\beta) \rangle$, $N = 3$ | $\langle P(\beta) \rangle$, $N = 4$ |
|---|---|---|
| 1.0 | 0.1671559(26) | 0.1251247(20) |
| 1.5 | 0.2536905(26) | 0.1884348(20) |
| 2.0 | 0.3504599(27) | 0.2540525(20) |
| 2.5 | 0.538850(11) | 0.3259904(23) |
| 3.0 | 0.6624336(27) | 0.4205164(39) |
| 3.5 | 0.7205217(22) | 0.6138239(25) |
| 4.0 | 0.7604072(18) | 0.6770339(19) |
| 4.5 | 0.7899299(20) | 0.7199132(19) |
| 5.0 | 0.8128151(16) | 0.7521754(15) |
| 5.5 | 0.8311878(14) | 0.7773942(16) |
| 6.0 | 0.8461740(18) | 0.7978683(12) |
| 6.5 | 0.8586757(17) | 0.8148922(11) |
| 7.0 | 0.8693439(14) | 0.8290391(13) |
| 7.5 | 0.8784713(11) | 0.8412828(11) |
| 8.0 | 0.8863930(16) | 0.8517954(18) |
| 8.5 | 0.8933310(17) | 0.86112031(93) |
| 9.0 | 0.8994785(12) | 0.8691295(10) |

$N = 4$. The plaquette susceptibility $\chi_p(\beta)$ was computed at each $\beta$. The results are presented in Fig. 6. At each volume, $\beta_c$ and $\chi_c$ were estimated using the multihistogram method. The obtained values are reported in Table VIII. No appreciable scaling of the peak values can be detected as $L$ is increased. Thus, from our data we can conclude that no phase transition is present for $N = 3$, 4, with the change of behavior from strong to weak coupling being described by a crossover. These conclusions are in agreement with the findings in Ref. [6] for the case $N = 3$.

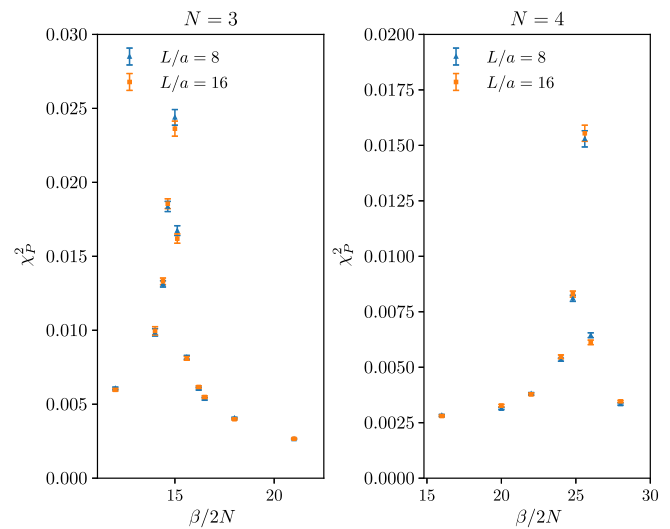

FIG. 6. The plaquette susceptibility $\chi_p$, as defined in Eq. (B2), as a function of $\beta = 2N/g^2$, at volumes $L/a = 8$ (orange squares), $L/a = 16$ (blue triangles) for the $Sp(2N)$ lattice gauge theory with $N = 3$ (left panel) and $N = 4$ (right panel).

---

[3]In principle, higher-order (e.g., third order) phase transitions are also possible. However, the only examples known to us arise strictly in the $N \to \infty$ limit (e.g., Refs. [141,142]). If a higher-order phase transition were present, it would be extremely difficult to detect it in our data. At the same time, we expect its influence on the numerical measurements to mimic a crossover. For this reason, we use here the expression *crossover* in a rather loose sense.

TABLE VIII. Location of the maximum value of the susceptibility for $N = 3$ and $N = 4$ at $L/a = 8, 12, 16$.

| | $N = 3$ | | $N = 4$ | |
|---|---|---|---|---|
| $L/a$ | $\beta_c$ | $\chi_{P\max}$ | $\beta_c$ | $\chi_{P\max}$ |
| 8 | 14.909(35) | 0.0319(14) | 24.45(3) | 0.045(6) |
| 12 | 14.909(41) | 0.0315(27) | 24.45(3) | 0.047(6) |
| 16 | 14.920(40) | 0.0283(87) | 24.45(3) | 0.048(7) |

Even if a phase transition is excluded, the presence of a crossover can still affect physical observables near the change of regime. An example of a similar effect in $SU(4)$ Yang-Mills with a fundamental Wilson action is described in Ref. [125], where the effect of the presence of a crossover reflects in a dip of the measured scalar glueball masses at the corresponding values of $\beta$. Similar results emerge also in $SU(2)$ with a mixed fundamental-adjoint action [143,144]. Therefore, to extrapolate lattice observables to the continuum limit with a simple and controlled dependence in $a\sqrt{\sigma}$, it is still necessary to be in the weak coupling regime. We achieved this by ensuring that our data points were far enough from the inflection points and then verifying that there was no visible signal of bulk phase contamination in our observables.

## APPENDIX C: CONTINUUM AND INFINITE VOLUME EXTRAPOLATIONS

As mentioned in Sec. III F, estimates of glueball and torelon masses obtained from the lattice are affected by systematic errors. We focus on the systematic error caused by working on a lattice of finite size in Sec. C 1 and on the error caused by the discretization in Sec. C 2.

### 1. Finite-size effects

The spectrum of a theory defined in a finite box of linear size $L$ with periodic boundary conditions depends on $L$. The problem was studied, for instance, in Ref. [145], and this dependence was found to be described by Eq. (47). The magnitude of the leading finite-size effects (FSEs) decays exponentially as a function of $mL$, where $m$ is the lightest excitation in the spectrum.

If $ma$ is estimated to a given finite precision, a value $L_{\min}/a$ exists such that for $L > L_{\min}$ the FSEs on the spectrum are negligible—by which we mean that their size is much smaller than the statistical error. For $L > L_{\min}$, the measured spectral masses can thus be considered as an estimate of the infinite-size spectrum at fixed lattice spacing. In the scaling regime, $mL$ is also a constant as $a \rightarrow 0$, and once $L_{\min}/a$ is found for a value of $a$, the FSEs will remain negligible as $a$ is taken to 0, provided the physical volume is kept approximately constant in the process. The precise value of $L_{\min}/a$ depends on the precision of our estimates and on the theory under scrutiny, and must be determined empirically.

To determine $L_{\min}$ and obtain an estimate for the spectrum at infinite size for $N = 3$, 4, we used the ensembles described in Sec. IV B. For each ensemble, we determined the glueball spectrum and the string tension. The results are reported in Table IX for $N = 3$ and Table X for $N = 4$. In this Appendix we focus on the lightest channel, which is consistently found to be $A_1^+$ and suffers from the largest FSEs. (Exceptions to this rule can be found, but they can only occur in the small $L/a$ regime, in which we are not interested.)

The estimates of $am_{A_1^+}$ are presented in Fig. 7, for $N = 3$ in the top panel and $N = 4$ in the bottom panel. From these figures we see that $am_{A_1^+}$ rapidly settles on a plateau as $L/a$ is increased. This means that, as expected, FSEs become negligible as $L$ is increased. A rough estimate yields $L_{\min}/a = 20$ for $Sp(6)$ at $\beta = 16.5$ and 10 for $Sp(8)$ at $\beta = 26.7$. As an *a posteriori* check, note that $m_{A_1^+}L_{\min} \sim 9.76$ for $Sp(6)$ and $m_{A_1^+}L_{\min} \sim 6.94$ for $Sp(8)$. The infinite-size spectrum can then be estimated by any one of the results at $L > L_{\min}$. We fitted Eq. (47) to the data using $b$ and $m(\infty)$ as fitting parameters. The fitted curves and the related $\chi^2/N_{\text{d.o.f.}}$ are displayed in Fig. 7.

TABLE IX. Estimates of the masses of glueballs $ma$ in all symmetry channels $R^P$ in units of the lattice spacing at $\beta = 16.5$, for various $L/a$ and for $N = 3$.

| | $L/a$ | | | | | |
|---|---|---|---|---|---|---|
| $R^P$ | 14 | 16 | 18 | 20 | 22 | 24 |
| $A_1^+$ | 0.321(14) | 0.492(21) | 0.419(29) | 0.488(19) | 0.479(18) | 0.493(19) |
| $A_1^-$ | 0.662(31) | 0.766(32) | 1.016(87) | 0.879(52) | 0.801(32) | 0.778(31) |
| $A_2^+$ | 1.122(75) | 1.125(88) | 1.172(78) | 1.108(93) | 1.282(92) | 1.087(66) |
| $A_2^-$ | 1.26(23) | 1.39(21) | 1.382(45) | 1.351(61) | 1.391(46) | 1.360(41) |
| $E^+$ | 0.374(19) | 0.496(27) | 0.684(32) | 0.754(28) | 0.755(25) | 0.731(14) |
| $E^-$ | 0.908(58) | 0.926(52) | 0.911(49) | 1.044(62) | 0.987(48) | 0.962(47) |
| $T_2^+$ | 0.726(34) | 0.782(32) | 0.748(30) | 0.684(30) | 0.775(28) | 0.740(22) |
| $T_2^-$ | 0.960(79) | 0.924(58) | 0.970(47) | 0.906(49) | 0.900(50) | 1.027(47) |
| $T_1^+$ | 1.19(11) | 1.16(11) | 1.154(98) | 1.091(88) | 1.20(12) | 1.196(26) |
| $T_1^-$ | 0.96(12) | 1.22(11) | 1.28(15) | 1.38(12) | 1.348(29) | 1.25(13) |

TABLE X. Estimates of the masses of glueballs $ma$ in all symmetry channels $R^P$ in units of the lattice spacing at $\beta = 26.7$ for various $L/a$ and for $N = 4$.

| $R^P$ | $L/a$ | | | | | | |
|---|---|---|---|---|---|---|---|
| | 8 | 10 | 12 | 14 | 16 | 18 | 20 |
| $A_1^+$ | 0.2881(60) | 0.694(19) | 0.695(31) | 0.664(29) | 0.727(33) | 0.685(27) | 0.707(25) |
| $A_1^-$ | 0.678(30) | 1.119(42) | 1.32(14) | 1.296(36) | 1.063(97) | 1.186(35) | 1.237(29) |
| $A_2^-$ | 2.05(25) | 1.98(22) | 1.62(16) | 1.91(23) | 2.05(19) | 2.05(20) | 2.55(31) |
| $A_2^+$ | 0.87(13) | 1.86(13) | 1.76(12) | 1.90(12) | 1.82(11) | 1.697(94) | 1.696(73) |
| $E^-$ | 0.776(63) | 1.544(70) | 1.37(22) | 1.446(68) | 1.518(64) | 1.543(47) | 1.473(50) |
| $E^+$ | 0.298(10) | 0.612(33) | 1.116(31) | 1.106(30) | 1.044(70) | 1.046(80) | 1.167(76) |
| $T_2^-$ | 1.13(20) | 1.377(62) | 1.564(79) | 1.505(64) | 0.97(14) | 1.503(64) | 1.412(46) |
| $T_2^+$ | 0.644(35) | 1.141(39) | 0.80(19) | 1.156(86) | 1.200(99) | 1.137(28) | 1.016(60) |
| $T_1^-$ | 1.83(19) | 2.08(25) | 2.09(19) | 1.99(20) | 2.09(20) | 2.18(18) | 2.12(15) |
| $T_1^+$ | 1.03(20) | 1.74(15) | 1.70(11) | 1.86(13) | 2.01(13) | 1.709(98) | 1.737(82) |

From the analysis above, we conclude that FSEs are negligible when $L > 20a$, for $N = 3$, and $L > 10a$, for $N = 4$. On these lattices, the condition $L\sqrt{\sigma} \geq 3$, which identifies the large volume regime of torelons, is also fulfilled. Hence, we choose this condition throughout as an indicator that finite volume effects can be neglected.

### 2. Continuum limit extrapolations

The behavior of the discretization error was studied in Sec. III F. The dimensionless ratio $m_{R^P}/\sqrt{\sigma}$ behaves, at leading order in $a$, as

$$\frac{m_{R^P}}{\sqrt{\sigma}}(a) = \frac{m_{R^P}}{\sqrt{\sigma}}(1 + c_{R^P}\sigma a^2), \tag{C1}$$

where $c_{R^P}$ is a constant that depends on the symmetry channel and on the excitation number. The multiplicative term on the right-hand side is the continuum limit of the ratio, while the term in $\sigma a^2$ describes the deviation with respect to this limit for sufficiently small $a$.

The continuum limit of the spectrum of glueballs can be obtained from sets of estimates thereof obtained at finite lattice spacing with a fit of Eq. (C1) to the data. This is the general strategy to extract results of the continuum spectrum from estimates at finite $a$, if the latter are obtained in a regime where Eq. (C1) is valid.

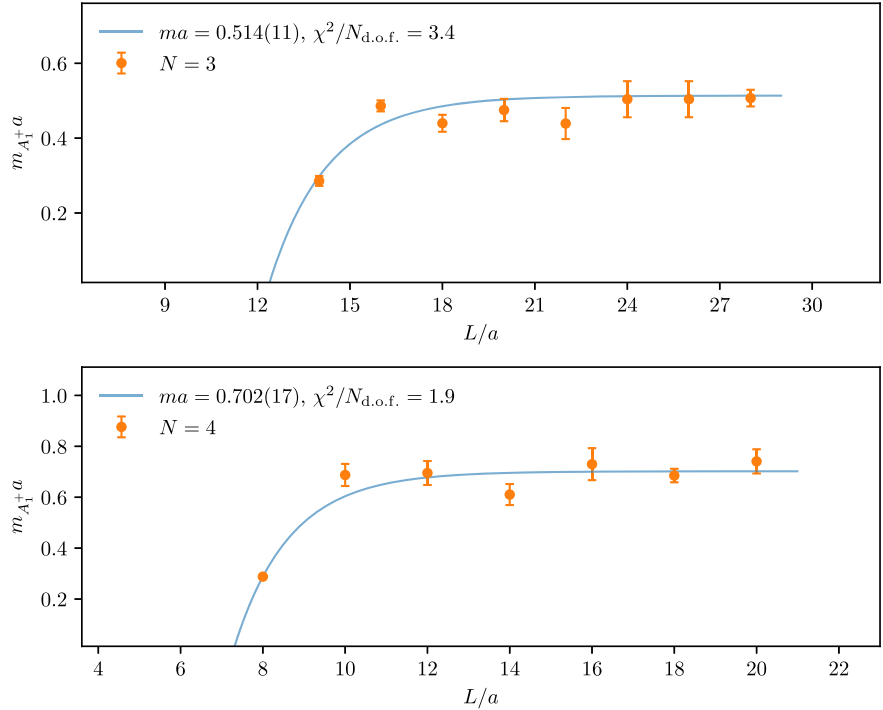

FIG. 7. Mass of the lightest glueball $m_{A_1^+}a$ in units of $a$, at fixed coupling, as a function of $L/a$. This corresponds to the $A_1^+$ channel for both $N = 3$, evaluated at $\beta = 16.5$ (top panel), and $N = 4$, evaluated at $\beta = 26.7$ (bottom panel). The solid line is the best fit of Eq. (47) to the data.

TABLE XI. Estimates of the glueball and the torelon masses for $N = 1$, in units of the lattice spacing $a$, on lattices of linear size $L$ and lattice spacing determined by the inverse coupling $\beta$. The error in brackets is discussed in the main text.

| $N = 1$ | $\beta = 2.2986$ $L = 10a$ | | $\beta = 2.3726$ $L = 12a$ | | $\beta = 2.4265$ $L = 16a$ | | $\beta = 2.5115$ $L = 20a$ | |
|---|---|---|---|---|---|---|---|---|
| | $m_{R^P}a$ | $\chi^2/N_{\text{d.o.f.}}$ | $m_{R^P}a$ | $\chi^2/N_{\text{d.o.f.}}$ | $m_{R^P}a$ | $\chi^2/N_{\text{d.o.f.}}$ | $m_{R^P}a$ | $\chi^2/N_{\text{d.o.f.}}$ |
| $A_1^+$ | 1.150(70) | $\cdots$ | 0.995(18) | 0.77 | 0.883(32) | 2.1 | 0.683(14) | 3.36 |
| $A_1^{+*}$ | 1.99(13) | 0.41 | 1.52(12) | $\cdots$ | 0.700(80) | $\cdots$ | 0.500(40) | $\cdots$ |
| $A_1^-$ | 2.11(23) | 1.02 | 1.53(30) | $\cdots$ | 1.430(80) | $\cdots$ | 1.030(60) | $\cdots$ |
| $A_1^{-*}$ | 2.33(20) | $\cdots$ | 2.45(45) | $\cdots$ | 1.83(24) | $\cdots$ | 1.30(20) | $\cdots$ |
| $A_2^+$ | 2.05(60) | $\cdots$ | 2.25(17) | 1.82 | 1.82(17) | $\cdots$ | 1.370(90) | $\cdots$ |
| $A_2^-$ | $\cdots$ | $\cdots$ | $\cdots$ | $\cdots$ | 1.99(20) | $\cdots$ | 1.50(20) | $\cdots$ |
| $T_2^+$ | 2.00(30) | $\cdots$ | 1.50(20) | $\cdots$ | 1.32(12) | 1.42 | 0.980(13) | 2.78 |
| $E^+$ | 2.00(50) | $\cdots$ | 1.24(25) | $\cdots$ | 1.229(98) | 1.99 | 0.950(60) | $\cdots$ |
| $T_2^-$ | 2.30(30) | $\cdots$ | 2.14(15) | 0.28 | 1.670(60) | $\cdots$ | 1.170(80) | $\cdots$ |
| $E^-$ | 2.10(40) | $\cdots$ | 2.07(14) | 0.64 | 1.59(14) | $\cdots$ | 1.220(60) | $\cdots$ |
| $T_1^+$ | $\cdots$ | $\cdots$ | 1.80(30) | $\cdots$ | 1.70(30) | $\cdots$ | 1.37(20) | $\cdots$ |
| $T_1^-$ | $\cdots$ | $\cdots$ | $\cdots$ | $\cdots$ | 2.00(20) | $\cdots$ | 1.50(15) | $\cdots$ |
| | $\sigma_s a^2$ | | $\sigma_s a^2$ | | $\sigma_s a^2$ | | $\sigma_s a^2$ | |
| | 0.1284(52) | $\cdots$ | 0.0736(31) | $\cdots$ | 0.0566(10) | $\cdots$ | 0.03116(63) | $\cdots$ |

TABLE XII. Estimates of glueball masses and string tensions for $N = 1$, in units of the lattice spacing $a$, on lattices of linear size $L$, and with lattice spacing determined by the inverse coupling $\beta$. The errors in brackets are discussed in the main text.

| $N = 1$ | $\beta = 2.6$ $L = 24a$ | | $\beta = 2.62$ $L = 26a$ | | $\beta = 2.7$ $L = 32a$ | |
|---|---|---|---|---|---|---|
| | $m_{R^P}a$ | $\chi^2/N_{\text{d.o.f.}}$ | $m_{R^P}a$ | $\chi^2/N_{\text{d.o.f.}}$ | $m_{R^P}a$ | $\chi^2/N_{\text{d.o.f.}}$ |
| $A_1^+$ | 0.467(24) | 2.16 | 0.487(32) | 2.82 | 0.356(10) | 0.71 |
| $A_1^{+*}$ | 0.97(11) | $\cdots$ | 0.680(50) | $\cdots$ | 1.390(90) | $\cdots$ |
| $A_1^-$ | 0.810(22) | 2.74 | 0.750(50) | $\cdots$ | 0.600(14) | 3.25 |
| $A_1^{-*}$ | 0.96(10) | $\cdots$ | 0.940(90) | $\cdots$ | 0.750(30) | $\cdots$ |
| $A_2^+$ | 0.900(90) | $\cdots$ | 0.896(28) | 2.81 | 0.680(50) | $\cdots$ |
| $A_2^-$ | 1.21(12) | $\cdots$ | 1.080(90) | $\cdots$ | 0.980(40) | $\cdots$ |
| $T_2^+$ | 0.690(50) | $\cdots$ | 0.660(40) | $\cdots$ | 0.490(30) | $\cdots$ |
| $E^+$ | 0.702(33) | 2.58 | 0.667(13) | 2.71 | 0.507(28) | 2.91 |
| $T_2^-$ | 0.900(50) | $\cdots$ | 0.830(50) | $\cdots$ | 0.687(65) | $\cdots$ |
| $E^-$ | 0.890(60) | $\cdots$ | 0.780(50) | $\cdots$ | 0.680(50) | $\cdots$ |
| $T_1^+$ | 1.050(80) | $\cdots$ | 1.000(70) | $\cdots$ | 0.820(40) | $\cdots$ |
| $T_1^-$ | 1.180(80) | $\cdots$ | 1.140(40) | $\cdots$ | 0.900(50) | $\cdots$ |
| | $\sigma_s a^2$ | | $\sigma_s a^2$ | | $\sigma_s a^2$ | |
| | 0.01715(26) | $\cdots$ | 0.01587(56) | $\cdots$ | 0.00938(18) | $\cdots$ |

For each $N = 1, 2, 3, 4$, ensembles of 10,000 thermalized configurations were obtained at several values of $a$ and $L/a$ and stored for later analysis. The values of $L/a$ were always chosen so that FSEs could be safely neglected. This has been verified *a posteriori* from the measured values of

$m_{R^P}L$. The glueball and torelon masses were estimated in units of the lattice spacing, as explained in Sec. III, for each ensemble.

While not strictly related to the continuum extrapolation, a comment is in order regarding the estimation of the uncertainty on $m_{R^P}a$ and to guide the reader in navigating the tables of results. To determine $t_{\text{min}}$, defined in Sec. III A, the quantity $m_{\text{eff}}(t)a$ is computed on all the available range of $t/a$. If a plateau can be found, the fits of Eq. (40) over the range $t > t_{\min}$ provide an estimate of $m_{R^P}a$ together with the *statistical error* of the fit and the corresponding $\chi^2/N_{\text{d.o.f.}}$. It is often the case, however, that the plateau is only $2a - 3a$ long, that an accurate determination of its preimage in $t/a$ is hindered by the contamination from larger mass states, or that the mass itself is large.

TABLE XIII. Estimates of glueball masses and string tensions for $N = 2$, in units of the lattice spacing $a$, on lattices of linear size $L$, and with lattice spacing determined by the inverse coupling $\beta$. The errors in brackets are discussed in the main text.

| $N = 2$ | $\beta = 7.62$ $L = 16a$ | | $\beta = 7.7$ $L = 16a$ | | $\beta = 7.85$ $L = 18a$ | | $\beta = 8.0$ $L = 20a$ | |
|---|---|---|---|---|---|---|---|---|
| | $m_{R^P}a$ | $\chi^2/N_{\text{d.o.f.}}$ | $m_{R^P}a$ | $\chi^2/N_{\text{d.o.f.}}$ | $m_{R^P}a$ | $\chi^2/N_{\text{d.o.f.}}$ | $m_{R^P}a$ | $\chi^2/N_{\text{d.o.f.}}$ |
| $A_1^+$ | 0.680(80) | ⋯ | 0.729(32) | 1.62 | 0.634(22) | 0.56 | 0.587(37) | 1.42 |
| $A_1^{+*}$ | ⋯ | ⋯ | 1.15(16) | ⋯ | 0.94(17) | ⋯ | 0.86(12) | ⋯ |
| $A_1^-$ | 1.21(20) | ⋯ | 1.190(50) | ⋯ | 0.980(60) | ⋯ | 0.880(40) | ⋯ |
| $A_1^{-*}$ | 1.57(32) | ⋯ | 1.64(26) | ⋯ | 1.39(12) | ⋯ | 1.230(80) | ⋯ |
| $A_2^+$ | 1.80(31) | ⋯ | 1.36(30) | ⋯ | 1.500(50) | ⋯ | 1.03(20) | ⋯ |
| $A_2^-$ | ⋯ | ⋯ | 1.85(30) | ⋯ | 1.40(30) | ⋯ | 1.38(20) | ⋯ |
| $T_2^+$ | ⋯ | ⋯ | 1.170(50) | ⋯ | 1.014(49) | 1.87 | 0.760(40) | ⋯ |
| $E^+$ | 0.96(24) | ⋯ | 1.160(28) | 1.86 | 0.910(50) | ⋯ | 0.810(50) | ⋯ |
| $T_2^-$ | ⋯ | ⋯ | 1.00(25) | ⋯ | 1.22(14) | ⋯ | 1.070(30) | ⋯ |
| $E^-$ | 1.30(35) | ⋯ | 1.24(20) | 1.58 | 1.16(13) | ⋯ | 1.060(60) | ⋯ |
| $T_1^+$ | ⋯ | ⋯ | 1.30(30) | ⋯ | 1.22(30) | ⋯ | 1.05(20) | ⋯ |
| $T_1^-$ | 1.60(40) | ⋯ | 2.07(17) | 0.56 | 1.58(17) | ⋯ | 1.10(20) | ⋯ |
| | $\sigma_s a^2$ | | $\sigma_s a^2$ | | $\sigma_s a^2$ | | $\sigma_s a^2$ | |
| | 0.0614(22) | ⋯ | 0.0517(12) | ⋯ | 0.03526(51) | ⋯ | 0.02487(66) | ⋯ |

TABLE XIV. Estimates of glueball masses and string tensions for $N = 2$, in units of the lattice spacing $a$, on lattices of linear size $L$, and with lattice spacing determined by the inverse coupling $\beta$. The errors in brackets are discussed in the main text.

| $N = 2$ | $\beta = 8.2$ $L = 26a$ | | $\beta = 8.3$ $L = 32a$ | |
|---|---|---|---|---|
| | $m_{R^P}a$ | $\chi^2/N_{\text{d.o.f.}}$ | $m_{R^P}a$ | $\chi^2/N_{\text{d.o.f.}}$ |
| $A_1^+$ | 0.445(21) | 2.31 | 0.402(12) | 1.57 |
| $A_1^{+*}$ | 0.710(80) | ⋯ | 0.640(50) | ⋯ |
| $A_1^-$ | 0.700(40) | ⋯ | 0.600(40) | ⋯ |
| $A_1^{-*}$ | 0.970(90) | ⋯ | 0.860(30) | ⋯ |
| $A_2^+$ | 1.000(50) | ⋯ | 0.880(70) | ⋯ |
| $A_2^-$ | 1.02(14) | ⋯ | 0.85(20) | ⋯ |
| $T_2^+$ | 0.610(50) | ⋯ | 0.570(50) | ⋯ |
| $E^+$ | 0.607(58) | 2.31 | 0.590(20) | ⋯ |
| $T_2^-$ | 0.820(60) | ⋯ | 0.740(60) | ⋯ |
| $E^-$ | 0.820(30) | ⋯ | 0.770(30) | ⋯ |
| $T_1^+$ | 1.000(50) | ⋯ | 0.790(80) | ⋯ |
| $T_1^-$ | 1.160(70) | ⋯ | 0.82(12) | ⋯ |
| | $\sigma_s a^2$ | | $\sigma_s a^2$ | |
| | 0.01676(26) | ⋯ | 0.01263(62) | ⋯ |

These difficulties in determining $t_{\min}$ lead to a systematic error on $m_{R^P}a$ that can be larger than the statistical error of the fitting procedure. In such cases, the statistical error of the fit cannot be trusted in describing the fluctuations of $m_{R^P}a$. Hence, we use a safe estimate of the mass and its error from the envelope of the points at plateau. A value for $\chi^2/N_{\text{d.o.f.}}$ cannot be defined, and the corresponding entry in the table is left empty. Finally, there are extreme cases for which an estimate for the mass simply cannot be found; i.e., a plateau is absent. In that case, the corresponding entry in the table is left empty.

All our estimates are reported in Tables XI–XX for the ensembles with $N = 1, 2, 3, 4$. The values of $\beta$ and $L/a$ are found in the first row of each table; the subsequent rows correspond to the ten symmetry channels, until the last row, which corresponds to the string tension. For each value of $N$, these estimates are plotted as a function of $\sigma a^2$ in Figs. 8–11.

In general, we found that the statistical errors and the confidence intervals are of the same order of magnitude, the latter being slightly larger in the majority of cases. This can be taken as an indication of the correctness of the method detailed above. In the following, we refer to the uncertainty in the determination of $m_{R^P}a$ generically as its "error."

A set of representative cases for the typical behavior of $m_{\text{eff}}(t)a$ is shown in Figs. 12 and 13, where the final estimate for $m_{R^P}a$ is represented as a dotted line and its error as the half-width of the corresponding horizontal band. In Fig. 12 the effective mass of state $A_1^+$ at the smallest available value of $a$ is reported for each $N = 1, 2, 3, 4$. For $N = 1, 2,$ and 3, a plateau can be identified. It is at least $t/a \sim 2$ long and starts at $t_{\min} = a$ for $N = 1, 3$ and

TABLE XV. Estimates of glueball masses and string tensions for $N = 3$, in units of the lattice spacing $a$, on lattices of linear size $L$, and with lattice spacing determined by the inverse coupling $\beta$. The errors in brackets are discussed in the main text.

| $N = 3$ | $\beta = 15.6$ $L = 12a$ | | $\beta = 15.65$ $L = 12a$ | | $\beta = 15.7$ $L = 12a$ | | $\beta = 15.85$ $L = 14a$ | |
|---|---|---|---|---|---|---|---|---|
| | $m_{R^P}a$ | $\chi^2/N_{\text{d.o.f.}}$ | $m_{R^P}a$ | $\chi^2/N_{\text{d.o.f.}}$ | $m_{R^P}a$ | $\chi^2/N_{\text{d.o.f.}}$ | $m_{R^P}a$ | $\chi^2/N_{\text{d.o.f.}}$ |
| $A_1^+$ | 0.765(27) | 0.18 | 0.777(26) | 0.07 | 0.750(24) | 0.56 | 0.720(20) | $\cdots$ |
| $A_1^{+*}$ | 1.43(16) | 0.99 | 1.29(12) | $\cdots$ | 1.167(96) | 1.28 | 1.17(12) | $\cdots$ |
| $A_1^-$ | 1.29(13) | 0.79 | 1.29(17) | 1.92 | 1.27(17) | $\cdots$ | 1.07(12) | $\cdots$ |
| $A_1^{-*}$ | 1.93(28) | $\cdots$ | 1.76(21) | $\cdots$ | 1.667(80) | 0.68 | 1.24(15) | $\cdots$ |
| $A_2^+$ | 1.80(15) | $\cdots$ | 1.92(11) | 0.51 | 1.64(54) | 1.94 | 1.54(20) | $\cdots$ |
| $A_2^-$ | 1.80(30) | $\cdots$ | 2.08(20) | $\cdots$ | 2.00(30) | $\cdots$ | 1.60(30) | $\cdots$ |
| $T_2^+$ | 1.06(15) | $\cdots$ | 1.23(12) | 2.05 | 1.213(87) | 2.94 | 1.075(55) | 0.49 |
| $E^+$ | 1.202(83) | 2.37 | 1.257(31) | 1.65 | 1.203(82) | 1.38 | 1.141(71) | 3.88 |
| $T_2^-$ | 1.50(13) | $\cdots$ | 1.70(14) | $\cdots$ | 1.40(20) | $\cdots$ | 1.37(14) | $\cdots$ |
| $E^-$ | 1.46(15) | $\cdots$ | 1.41(30) | $\cdots$ | 1.33(18) | $\cdots$ | 1.26(12) | $\cdots$ |
| $T_1^+$ | 1.70(30) | $\cdots$ | 2.00(18) | $\cdots$ | 1.07(40) | $\cdots$ | 1.52(16) | $\cdots$ |
| $T_1^-$ | 1.37(35) | $\cdots$ | 1.90(40) | $\cdots$ | 2.25(35) | $\cdots$ | 1.70(40) | $\cdots$ |
| | $\sigma_s a^2$ | | $\sigma_s a^2$ | | $\sigma_s a^2$ | | $\sigma_s a^2$ | |
| | 0.06464(79) | $\cdots$ | 0.0663(11) | $\cdots$ | 0.05918(72) | $\cdots$ | 0.04879(60) | $\cdots$ |

TABLE XVI. Estimates of glueball masses and string tensions for $N = 3$, in units of the lattice spacing $a$, on lattices of linear size $L$, and with lattice spacing determined by the inverse coupling $\beta$. The errors in brackets are discussed in the main text.

| $N = 3$ | $\beta = 16.1$ $L = 16a$ | | $\beta = 16.3$ $L = 20a$ | | $\beta = 16.5$ $L = 20a$ | | $\beta = 16.7$ $L = 28a$ | |
|---|---|---|---|---|---|---|---|---|
| | $m_{R^P}a$ | $\chi^2/N_{\text{d.o.f.}}$ | $m_{R^P}a$ | $\chi^2/N_{\text{d.o.f.}}$ | $m_{R^P}a$ | $\chi^2/N_{\text{d.o.f.}}$ | $m_{R^P}a$ | $\chi^2/N_{\text{d.o.f.}}$ |
| $A_1^+$ | 0.581(28) | 0.25 | 0.536(20) | 2.09 | 0.499(15) | 0.11 | 0.432(10) | $\cdots$ |
| $A_1^{+*}$ | 0.95(12) | $\cdots$ | 0.910(80) | $\cdots$ | 0.810(60) | $\cdots$ | 0.690(90) | $\cdots$ |
| $A_1^-$ | 0.971(41) | 0.49 | 0.846(36) | 0.33 | 0.808(21) | 2.62 | 0.610(50) | $\cdots$ |
| $A_1^{-*}$ | 1.16(17) | $\cdots$ | 1.031(53) | 1.78 | 0.82(17) | $\cdots$ | 0.72(16) | $\cdots$ |
| $A_2^+$ | 1.488(38) | 3.97 | 1.23(15) | $\cdots$ | 1.050(50) | $\cdots$ | 0.92(12) | $\cdots$ |
| $A_2^-$ | 1.48(25) | $\cdots$ | 1.50(12) | $\cdots$ | 1.28(10) | $\cdots$ | 1.12(10) | $\cdots$ |
| $T_2^+$ | 0.854(85) | 1.14 | 0.780(60) | $\cdots$ | 0.700(40) | $\cdots$ | 0.636(20) | 2.92 |
| $E^+$ | 0.954(35) | 0.98 | 0.830(50) | $\cdots$ | 0.710(30) | $\cdots$ | 0.650(30) | $\cdots$ |
| $T_2^-$ | 1.200(50) | $\cdots$ | 1.080(81) | 3.73 | 0.950(60) | $\cdots$ | 0.770(60) | $\cdots$ |
| $E^-$ | 1.247(24) | 1.94 | 1.090(70) | $\cdots$ | 0.974(37) | 0.33 | 0.830(50) | $\cdots$ |
| $T_1^+$ | 1.50(10) | $\cdots$ | 1.380(90) | $\cdots$ | 1.12(12) | $\cdots$ | 0.920(80) | $\cdots$ |
| $T_1^-$ | 1.59(10) | $\cdots$ | 1.20(16) | $\cdots$ | 1.10(20) | $\cdots$ | 1.21(10) | $\cdots$ |
| | $\sigma_s a^2$ | | $\sigma_s a^2$ | | $\sigma_s a^2$ | | $\sigma_s a^2$ | |
| | 0.03501(59) | $\cdots$ | 0.02825(99) | $\cdots$ | 0.02303(33) | $\cdots$ | 0.01606(33) | $\cdots$ |

$t_{\min} = 2a$ for $N = 2$. An estimate for $m_{R^P}a$ can thus be obtained from a fit of Eq. (40) over the range $t_{\min} < t < t_{\max}$. The value $t_{\max}$ was chosen to be as large as possible while still resulting in a value of $\chi^2/N_{\text{d.o.f.}}$ close to 1 for the corresponding fit. In most cases of the analysis where a plateau could be identified, and in particular in all of those depicted in Fig. 12, it was possible to set $t_{\max} = La/2$. For $N = 4$, to the contrary, $m_{A_1^+}a$ is estimated from the envelope of the quasiplateau that starts at $t/a \sim 1$ and that is only $a$ long. In Fig. 13, the effective mass of state $T_1^+$ at $N = 3$ is reported for a range of lattice spacings, from coupling $\beta = 15.85$ to coupling $\beta = 17.1$. In all of these cases a plateau of length $2 - 3a$ cannot be found and $m_{T_1^+}a$ is estimated from the envelope of a quasiplateau. At $\beta = 17.1$, the contamination of higher mass states lasts up to $t/a = 4$, and a quasiplateau can only be identified for $t/a$ in the interval $\sim [3a, 5a]$. Its quality progressively degrades as $\beta$ takes smaller values, becoming barely visible for $\beta = 15.85$.

TABLE XVII. Estimates of glueball masses and string tensions for $N = 3$, in units of the lattice spacing $a$, on lattices of linear size $L$, and with lattice spacing determined by the inverse coupling $\beta$. The errors in brackets are discussed in the main text.

| | $\beta = 16.8$ $L = 24a$ | | $\beta = 17.1$ $L = 28a$ | |
|---|---|---|---|---|
| $N = 3$ | $m_{R^P}a$ | $\chi^2/N_{\text{d.o.f.}}$ | $m_{R^P}a$ | $\chi^2/N_{\text{d.o.f.}}$ |
| $A_1^+$ | 0.441(15) | $\cdots$ | 0.360(14) | 1.02 |
| $A_1^{+*}$ | 0.730(70) | $\cdots$ | 0.610(60) | $\cdots$ |
| $A_1^-$ | 0.66(12) | $\cdots$ | 0.550(30) | $\cdots$ |
| $A_1^{-*}$ | 1.49(10) | $\cdots$ | 0.970(90) | $\cdots$ |
| $A_2^+$ | 0.76(28) | $\cdots$ | 0.764(77) | 2.14 |
| $A_2^-$ | 0.90(17) | $\cdots$ | 0.99(13) | $\cdots$ |
| $T_2^+$ | 0.680(30) | $\cdots$ | 0.558(19) | 2.01 |
| $E^+$ | 0.663(22) | 1.93 | 0.560(18) | 1.17 |
| $T_2^-$ | 0.840(70) | $\cdots$ | 0.730(40) | $\cdots$ |
| $E^-$ | 0.787(79) | 3.98 | 0.690(50) | $\cdots$ |
| $T_1^+$ | 0.80(30) | $\cdots$ | 0.73(14) | $\cdots$ |
| $T_1^-$ | 1.16(10) | $\cdots$ | 0.85(15) | $\cdots$ |
| | $\sigma_s a^2$ | | $\sigma_s a^2$ | |
| | 0.01824(30) | $\cdots$ | 0.01183(52) | $\cdots$ |

In general, a plateau can be identified when $a$ is sufficiently small that the mass of the ground state in lattice units is well below $1/a$. In principle, hierarchies between the masses in the spectrum may make their estimation difficult for all the channels at a common value of $a$. This is shown for channels $A_1^+$ and $T_1^+$ at $\beta = 17.1$ in

Figs. 12 and 13, respectively, where the mass of the $T_1^+$ is approximately twice the mass of $A_1^+$. Moreover, as discussed in Sec. III F, the plateau may show contaminations from excited and scattering states. These may be difficult to remove even for small $a$, as shown for $T_1^+$ in Fig. 13. Despite these difficulties, it was possible to provide reliable estimates of $m_{R^P}a$ in a great majority of cases.

Let us now comment on the features of these finite-$a$ estimates that are common across all the values of $N$. The fact that in every case $m_{R^P}L > 3$ allows us to safely neglect FSEs for all the ensembles, as anticipated. Moreover, all the estimates satisfy $m_{R^P}a \leq 2$, except for the two roughest lattices when $N = 1$, corresponding to $\beta = 2.2986$ and $\beta = 2.3726$. Therefore, we can assert that our choices of $\beta$ are well calibrated to study the flow to the continuum limit of the spectrum of these systems. In the glueball sector, the $A_1^+$ is consistently the lightest channel, followed by the $(E^{\pm}, T_2^{\pm})$ degenerate pairs. The error of the estimates is larger for larger $m_{R^P}a$, as is to be expected on the basis of the discussion in Sec. III F. The $E^{\pm}$ and $T_2^{\pm}$ are degenerate over the whole range of $a$ probed at least at the $2\sigma$ level, with the mass difference being below 1 standard deviation in most of the cases. This is a nontrivial a posteriori test of the restoration of continuum rotational invariance and can be taken as a signal that we are in the regime for which Eq. (48) is valid.

An additional source of systematic error, the effects of which are difficult to account for, is the autocorrelation time of the system, which grows as the continuum limit is approached. Since the topological charge is one of the quantities with the longest autocorrelation time, studying the evolution of this observable yields a conservative estimator of these effects. Particular attention should be paid to cases in which the topology is (nearly) frozen. These can be detected by analyzing the time series of the topological charge. To this end, a subset of configurations were obtained from the $N = 3$ and $N = 4$ ensembles, by picking one configuration each 100. The gradient flow was then used to smooth each of the configurations, and the *regularized* topological charge was computed using Eq. (11) at each smoothing step. The continuum topological charge is obtained when the regularized topological charge reaches a plateau under further smoothing operations. The results of this analysis are visible in Figs. 14 and 15.

TABLE XVIII. Estimates of glueball masses and string tensions for $N = 4$, in units of the lattice spacing $a$, on lattices of linear size $L$, and with lattice spacing determined by the inverse coupling $\beta$. The errors in brackets are discussed in the main text.

| | $\beta = 26.5$ $L = 14a$ | | $\beta = 26.7$ $L = 14a$ | | $\beta = 26.8$ $L = 14a$ | | $\beta = 27.0$ $L = 16a$ | |
|---|---|---|---|---|---|---|---|---|
| $N = 4$ | $m_{R^P}a$ | $\chi^2/N_{\text{d.o.f.}}$ | $m_{R^P}a$ | $\chi^2/N_{\text{d.o.f.}}$ | $m_{R^P}a$ | $\chi^2/N_{\text{d.o.f.}}$ | $m_{R^P}a$ | $\chi^2/N_{\text{d.o.f.}}$ |
| $A_1^+$ | 0.705(33) | 2.88 | 0.734(12) | 0.6 | 0.705(22) | 0.26 | 0.615(30) | 0.08 |
| $A_1^{+*}$ | 1.22(12) | $\cdots$ | 1.262(88) | 1.08 | 1.104(66) | 2.13 | 0.94(11) | $\cdots$ |
| $A_1^-$ | 1.230(80) | $\cdots$ | 1.198(28) | 1.06 | 1.140(60) | $\cdots$ | 1.055(59) | 0.53 |
| $A_1^{-*}$ | 1.73(26) | $\cdots$ | 1.66(15) | $\cdots$ | 1.564(48) | 1.51 | 0.900(90) | $\cdots$ |
| $A_2^+$ | 1.890(50) | $\cdots$ | 1.21(25) | 1.97 | 1.720(60) | $\cdots$ | 1.500(50) | $\cdots$ |
| $A_2^-$ | 2.03(12) | 2.56 | 1.99(30) | $\cdots$ | 2.06(15) | 3.09 | 1.980(90) | $\cdots$ |
| $T_2^+$ | 1.15(15) | $\cdots$ | 1.081(67) | 0.23 | 1.048(53) | 1.38 | 1.028(15) | 2.08 |
| $E^+$ | 1.15(14) | $\cdots$ | 1.156(69) | 1.3 | 1.210(59) | 12.53 | 1.010(40) | $\cdots$ |
| $T_2^-$ | 1.60(10) | $\cdots$ | 1.46(15) | $\cdots$ | 1.370(40) | $\cdots$ | 1.310(40) | $\cdots$ |
| $E^-$ | 1.620(40) | $\cdots$ | 1.40(10) | $\cdots$ | 1.26(13) | 1.07 | 1.310(60) | $\cdots$ |
| $T_1^+$ | 1.40(30) | $\cdots$ | 1.50(15) | $\cdots$ | 1.690(70) | $\cdots$ | 1.590(90) | $\cdots$ |
| $T_1^-$ | 1.50(30) | $\cdots$ | 1.95(15) | $\cdots$ | 1.94(10) | $\cdots$ | 1.750(80) | $\cdots$ |
| | $\sigma_s a^2$ | | $\sigma_s a^2$ | | $\sigma_s a^2$ | | $\sigma_s a^2$ | |
| | 0.06386(85) | $\cdots$ | 0.0549(15) | $\cdots$ | 0.04947(72) | $\cdots$ | 0.04362(61) | $\cdots$ |

TABLE XIX.   Estimates of glueball masses and string tensions for $N = 4$, in units of the lattice spacing $a$, on lattices of linear size $L$, and with lattice spacing determined by the inverse coupling $\beta$. The errors in brackets are discussed in the main text.

| $N = 4$ | $\beta = 27.2$ $L = 16a$ | | $\beta = 27.3$ $L = 16a$ | | $\beta = 27.6$ $L = 18a$ | | $\beta = 27.9$ $L = 20a$ | |
| --- | --- | --- | --- | --- | --- | --- | --- | --- |
| | $m_{R^P}a$ | $\chi^2/N_{\text{d.o.f.}}$ | $m_{R^P}a$ | $\chi^2/N_{\text{d.o.f.}}$ | $m_{R^P}a$ | $\chi^2/N_{\text{d.o.f.}}$ | $m_{R^P}a$ | $\chi^2/N_{\text{d.o.f.}}$ |
| $A_1^+$ | 0.610(20) | $\cdots$ | 0.565(20) | 1.96 | 0.530(20) | $\cdots$ | 0.486(18) | $\cdots$ |
| $A_1^{+*}$ | 0.91(10) | $\cdots$ | 0.860(60) | $\cdots$ | 0.760(50) | $\cdots$ | 1.27(15) | $\cdots$ |
| $A_1^-$ | 1.025(51) | 2.16 | 0.890(90) | $\cdots$ | 0.890(30) | $\cdots$ | 0.769(33) | 1.79 |
| $A_1^{-*}$ | 1.32(17) | $\cdots$ | 1.32(16) | $\cdots$ | 1.30(12) | $\cdots$ | 1.11(14) | $\cdots$ |
| $A_2^+$ | 1.480(70) | $\cdots$ | 1.410(80) | $\cdots$ | 1.230(90) | $\cdots$ | 1.164(84) | 1.3 |
| $A_2^-$ | 1.67(15) | $\cdots$ | 1.690(70) | $\cdots$ | 1.45(10) | $\cdots$ | 1.23(12) | $\cdots$ |
| $T_2^+$ | 0.946(41) | 0.98 | 0.863(49) | 0.67 | 0.815(26) | 1.84 | 0.700(50) | $\cdots$ |
| $E^+$ | 0.957(77) | 1.94 | 0.870(50) | $\cdots$ | 0.839(27) | 0.41 | 0.690(60) | $\cdots$ |
| $T_2^-$ | 1.200(60) | $\cdots$ | 1.160(60) | $\cdots$ | 1.100(90) | $\cdots$ | 0.972(63) | 1.97 |
| $E^-$ | 1.220(50) | $\cdots$ | 1.168(36) | 2.59 | 1.030(60) | $\cdots$ | 0.983(57) | 0.94 |
| $T_1^+$ | 1.41(16) | 1.36 | 1.480(90) | $\cdots$ | 1.290(90) | $\cdots$ | 1.230(50) | $\cdots$ |
| $T_1^-$ | 1.60(10) | $\cdots$ | 1.45(15) | $\cdots$ | 1.48(15) | $\cdots$ | 1.33(10) | $\cdots$ |
| | $\sigma_s a^2$ | | $\sigma_s a^2$ | | $\sigma_s a^2$ | | $\sigma_s a^2$ | |
| | 0.03644(56) | $\cdots$ | 0.03384(56) | $\cdots$ | 0.02728(48) | $\cdots$ | 0.02303(54) | $\cdots$ |

TABLE XX.   Estimates of glueball masses and string tensions for $N = 4$, in units of the lattice spacing $a$, on lattices of linear size $L$, and with lattice spacing determined by the inverse coupling $\beta$. The errors in brackets are discussed in the main text.

| $N = 4$ | $\beta = 28.3$ $L = 22a$ | |
| --- | --- | --- |
| | $m_{R^P}a$ | $\chi^2/N_{\text{d.o.f.}}$ |
| $A_1^+$ | 0.440(10) | $\cdots$ |
| $A_1^{+*}$ | 0.910(50) | $\cdots$ |
| $A_1^-$ | 0.680(70) | $\cdots$ |
| $A_1^{-*}$ | 1.050(90) | $\cdots$ |
| $A_2^+$ | 0.940(80) | $\cdots$ |
| $A_2^-$ | 1.14(11) | $\cdots$ |
| $T_2^+$ | 0.641(22) | 0.64 |
| $E^+$ | 0.634(35) | 1.78 |
| $T_2^-$ | 0.849(38) | 2.91 |
| $E^-$ | 0.850(60) | $\cdots$ |
| $T_1^+$ | 1.020(80) | $\cdots$ |
| $T_1^-$ | 1.220(90) | $\cdots$ |
| | $\sigma_s a^2$ | |
| | 0.01869(50) | $\cdots$ |

For both $N = 3$ and $N = 4$, there is a value $\beta_{\min}$ above which the topological charge barely changes with the Monte Carlo steps that we are able to perform. These ensembles are topologically frozen. From visual inspection of the figures we estimate that

$$\beta_{\min}(N = 3) \simeq 16.5, \qquad \beta_{\min}(N = 4) \simeq 27.0. \quad \text{(C2)}$$

Given that topological freezing affects only our two largest values of $N$, where the systematic effects it induces on measurements of masses are expected to become less severe (as discussed in Sec. III F), we included the estimates obtained from these frozen ensembles in the extrapolation to the continuum limit.

A related potential source of systematic error lies in the length of the initial thermalization. Our earlier estimates of the continuum spectrum, especially for $N = 3$ and $N = 4$, presented a visible dip in the calculated masses for the smallest values of $\sigma a^2$. This urged us to perform an overall check of the invariance of the final result under the prolongation of the simulation trajectory. In Fig. 16, we show results of $m_{R^P}/\sqrt{\sigma}$ at finite lattice spacing as a function of the initial thermalization time. The fact that these estimates are largely independent of this initial thermalization time suggests that the Markov chains from which the final averages are obtained are long enough for the system to be at statistical equilibrium.

Let us now discuss the continuum extrapolations of the ratios $m_{R^P}/\sqrt{\sigma}$ for given values of $a$. These ratios can easily be formed for each ensemble from the estimates in Tables XI–XX. At each value of $N$, fits of Eq. (C1) using

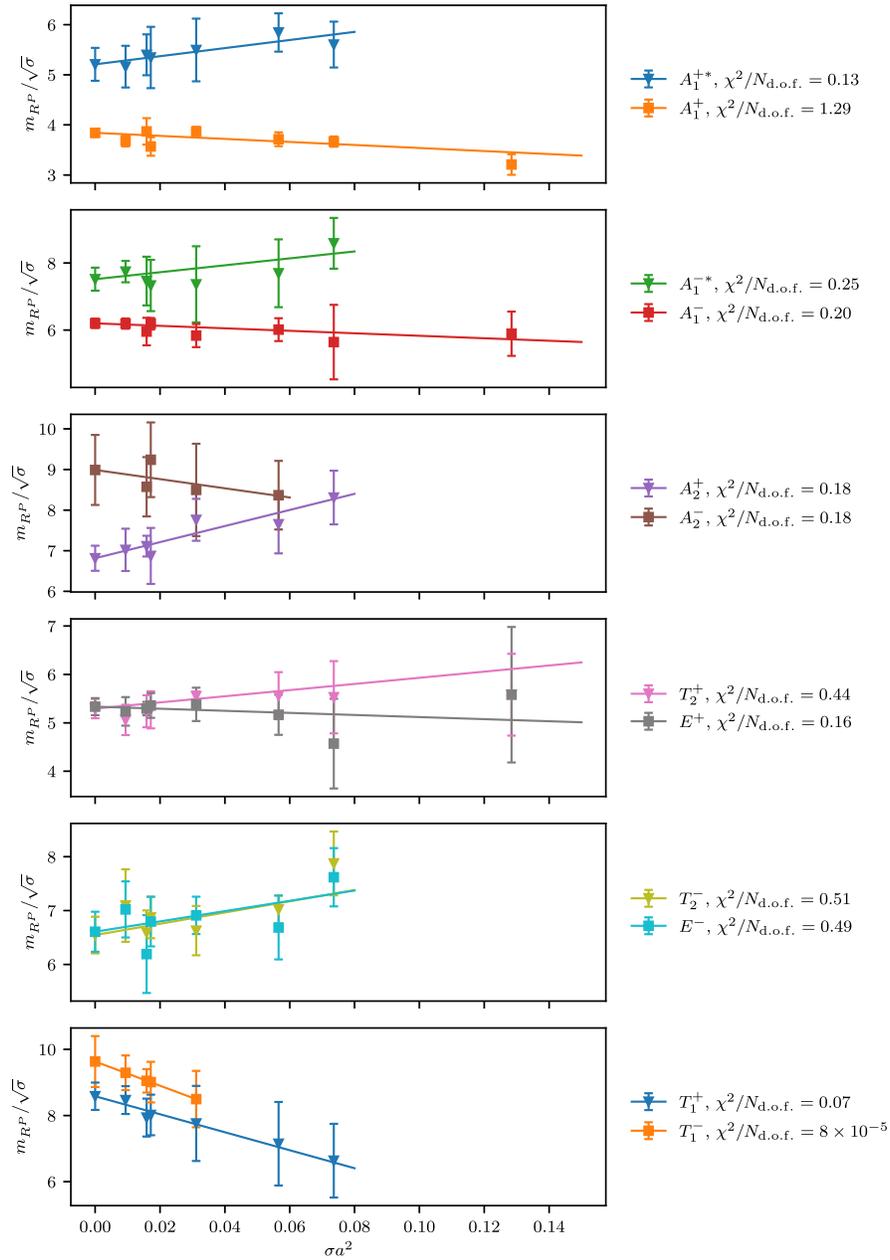

FIG. 8. Glueball mass in each symmetry channel $R^P$ of the $Sp(2N)$ theory with $N = 1$, in units of $\sqrt{\sigma}$, as a function of $\sigma a^2$. For each symmetry channel $R^P$, the value at $\sigma a^2 = 0$ is the continuum limit, obtained from the best fit of Eq. (C1) to the data. The best fits lines are represented as solid lines.

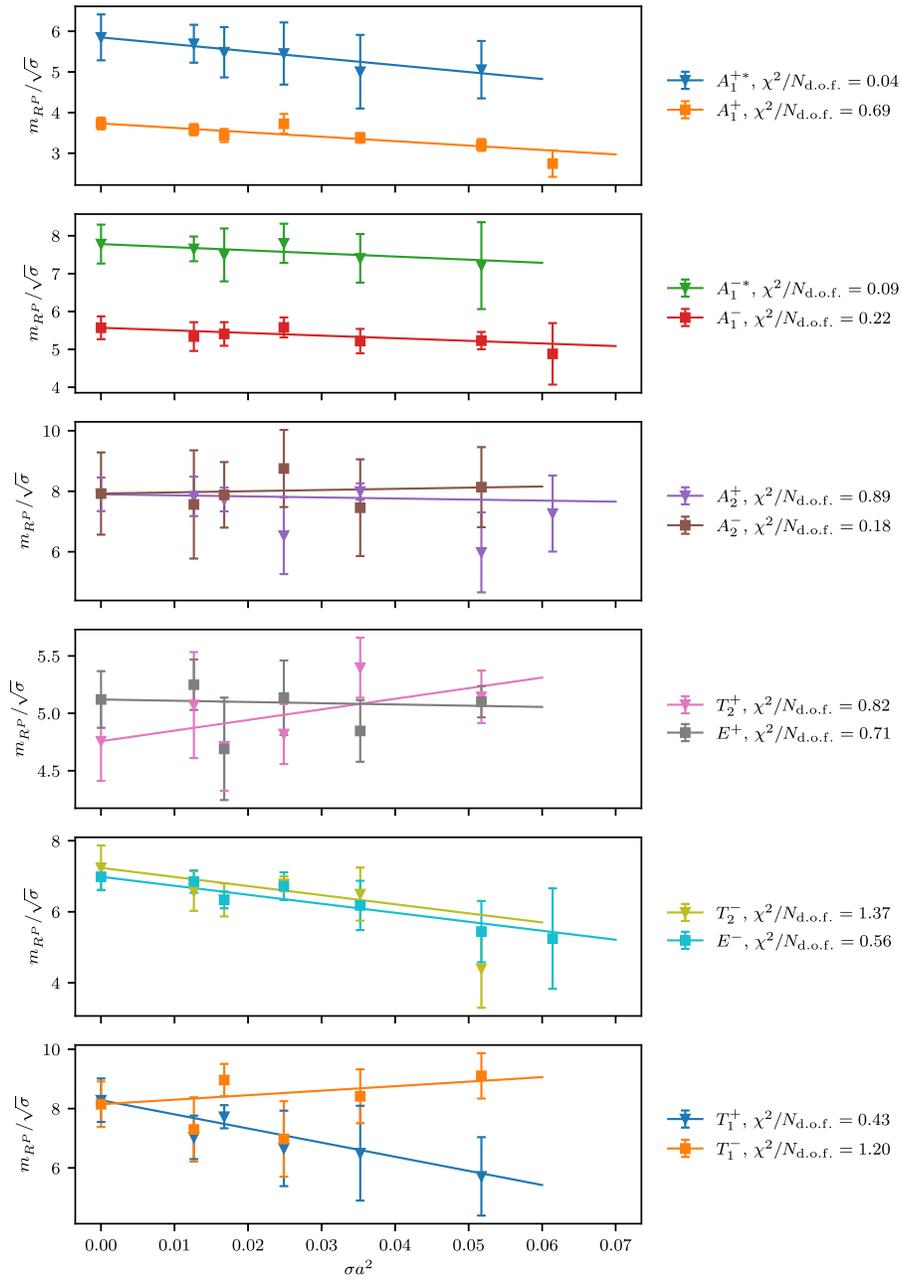

FIG. 9. Glueball mass in each symmetry channel $R^P$ of the $Sp(2N)$ theory with $N = 2$, in units of $\sqrt{\sigma}$, as a function of $\sigma a^2$. For each symmetry channel $R^P$, the value at $\sigma a^2 = 0$ is the continuum limit, obtained from the best fit of Eq. (C1) to the data. The best fits lines are represented as solid lines.

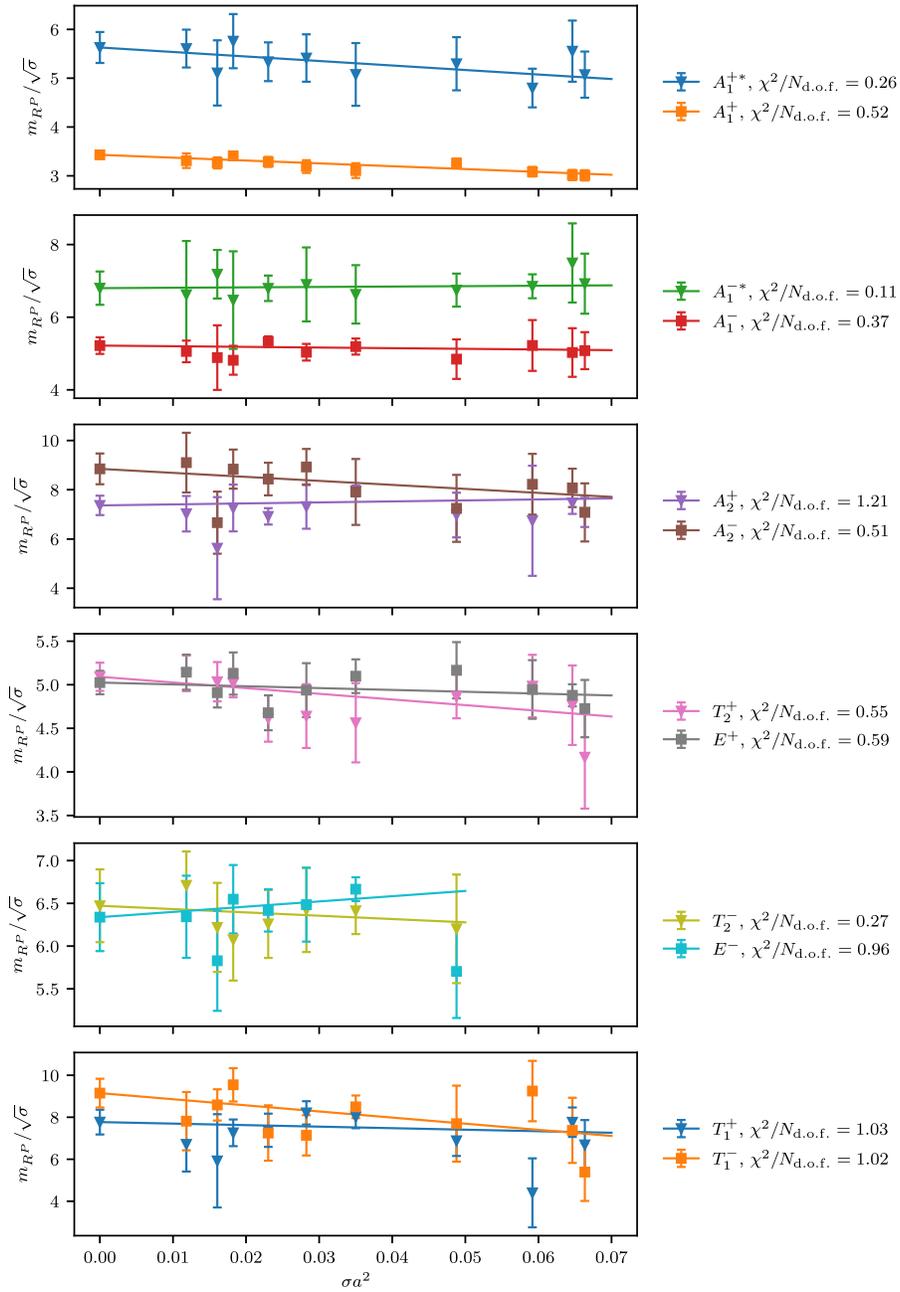

FIG. 10. Glueball mass in each symmetry channel $R^P$ of the $Sp(2N)$ theory with $N = 3$, in units of $\sqrt{\sigma}$, as a function of $\sigma a^2$. For each symmetry channel $R^P$, the value at $\sigma a^2 = 0$ is the continuum limit, obtained from the best fit of Eq. (C1) to the data. The best fits lines are represented as solid lines.

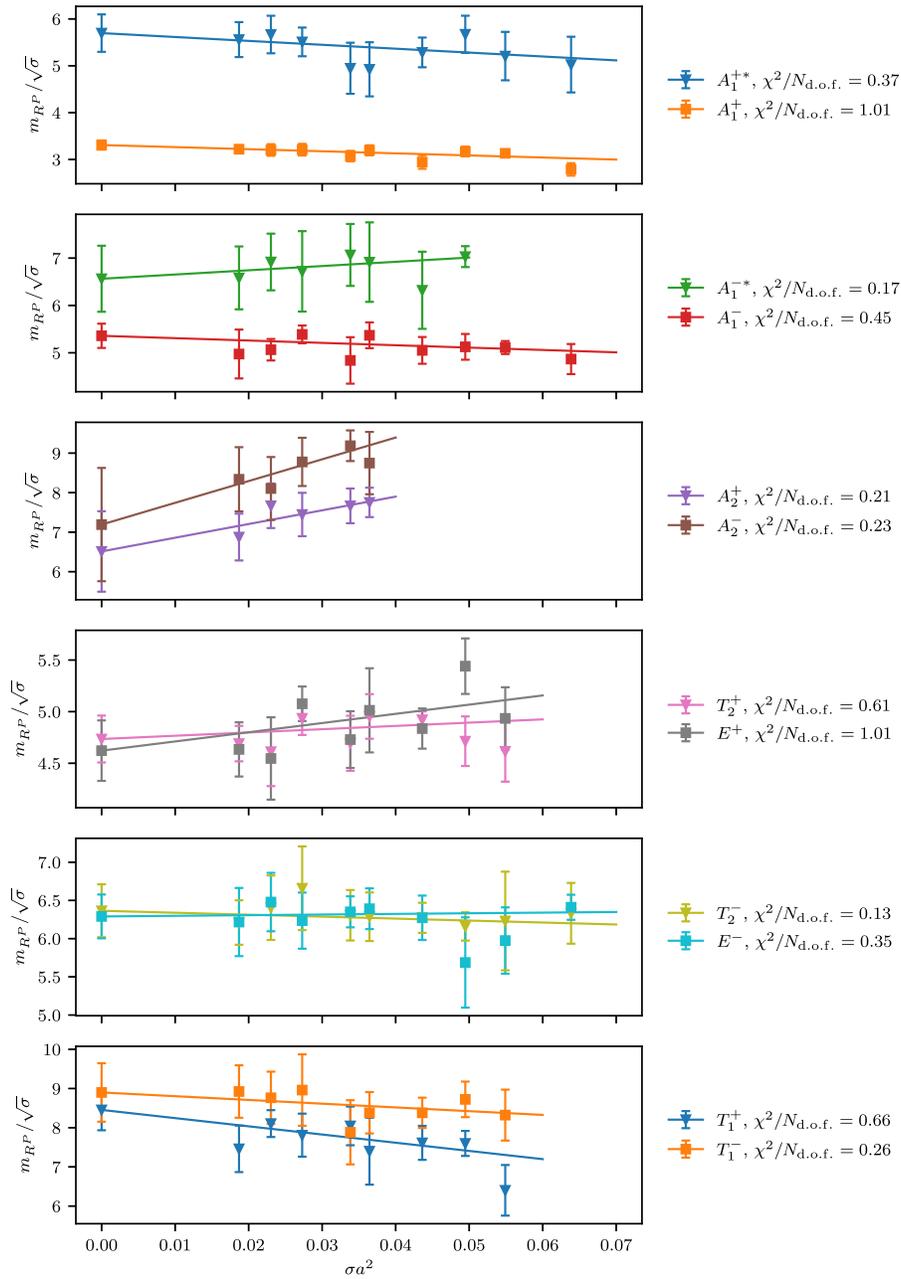

FIG. 11.  Glueball mass in each symmetry channel $R^P$ of the $Sp(2N)$ theory with $N = 4$, in units of $\sqrt{\sigma}$, as a function of $\sigma a^2$. For each symmetry channel $R^P$, the value at $\sigma a^2 = 0$ is the continuum limit, obtained from the best fit of Eq. (C1) to the data. The best fits lines are represented as solid lines.

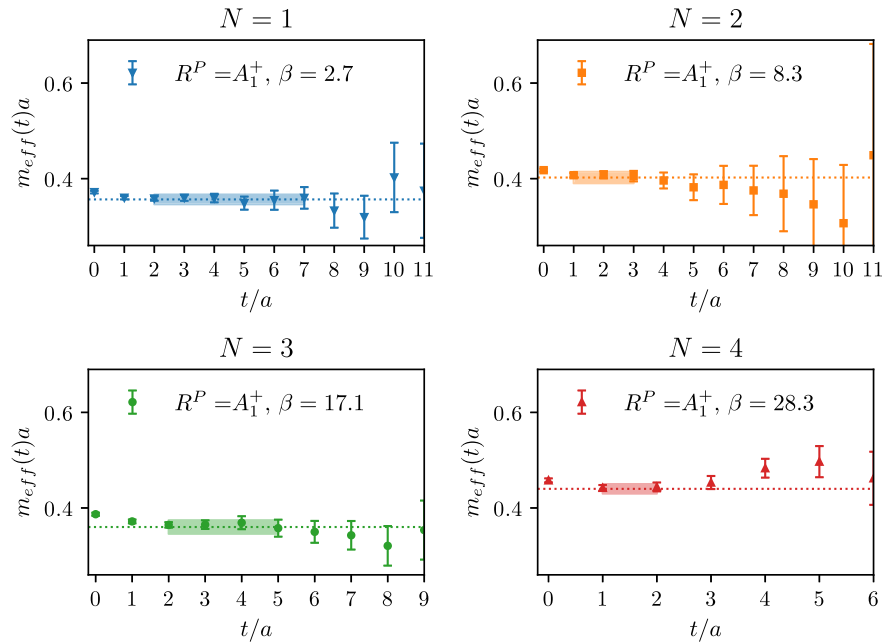

FIG. 12. The effective mass $m_{\text{eff}}(t)a$ for the channel $R^P = A_1^+$, as a function of $t/a$, obtained at the smallest available lattice spacing for each value of $N = 1, 2, 3, 4$. The plateau is denoted by a band. Its horizontal extent covers the $t$ interval of the plateau, and its half vertical width is the error of the final estimate for $m_{A_1^+}a$, which is represented as a dotted horizontal line. In this case, a plateau can be identified for all values of $N$ other than $N = 4$, which corresponds to the largest value of $m_{\text{eff}}(t)a$ among those shown in the figure.

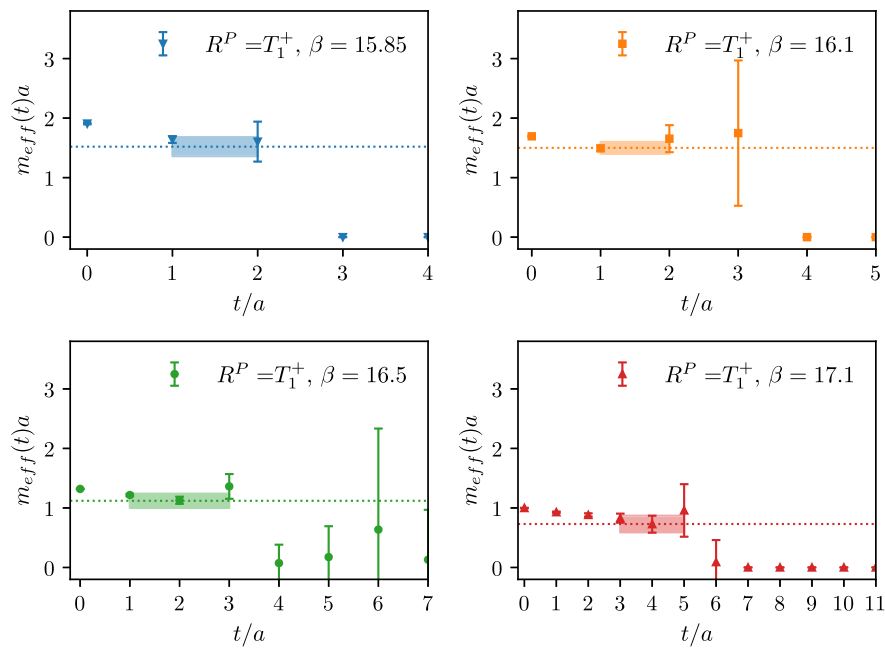

FIG. 13. The effective mass $m_{\text{eff}}(t)a$ for the channel $R^P = T_1^+$, as a function of $t/a$, for $N = 3$, obtained on a set of lattice spacings corresponding to couplings $\beta = 15.85$, $\beta = 16.1$, $\beta = 16.5$, and $\beta = 17.1$. In this case, only *quasiplateaux* could be identified, as a consequence of the large mass of the $T_1^+$ state in lattice units and of contaminations from excited and scattering states. The quasiplateau is denoted by a band. Its horizontal extent covers the $t$ interval of the quasiplateau, and its half vertical width is the error of the final estimate for $m_{T_1^+}a$, which is represented as a dotted horizontal line. A similar behavior is observed for the other values of $N$.

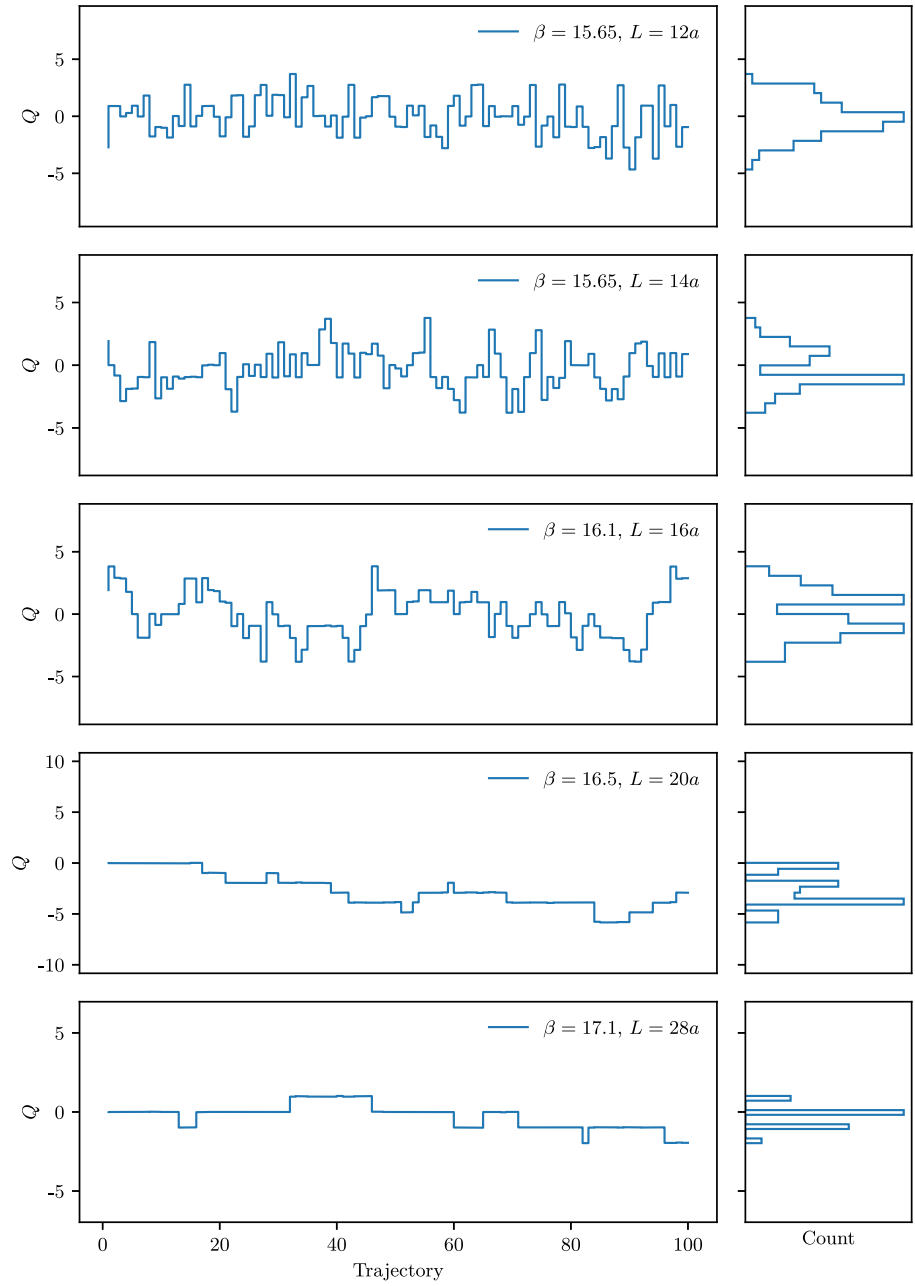

FIG. 14. Histories and statistical distributions (histograms) of the topological charge defined in Eq. (11) for the ensembles obtained at $N = 3$. The configurations of each ensemble are smoothed with the gradient flow defined in Eq. (16). The frequency of sampling in running time is described in the text.

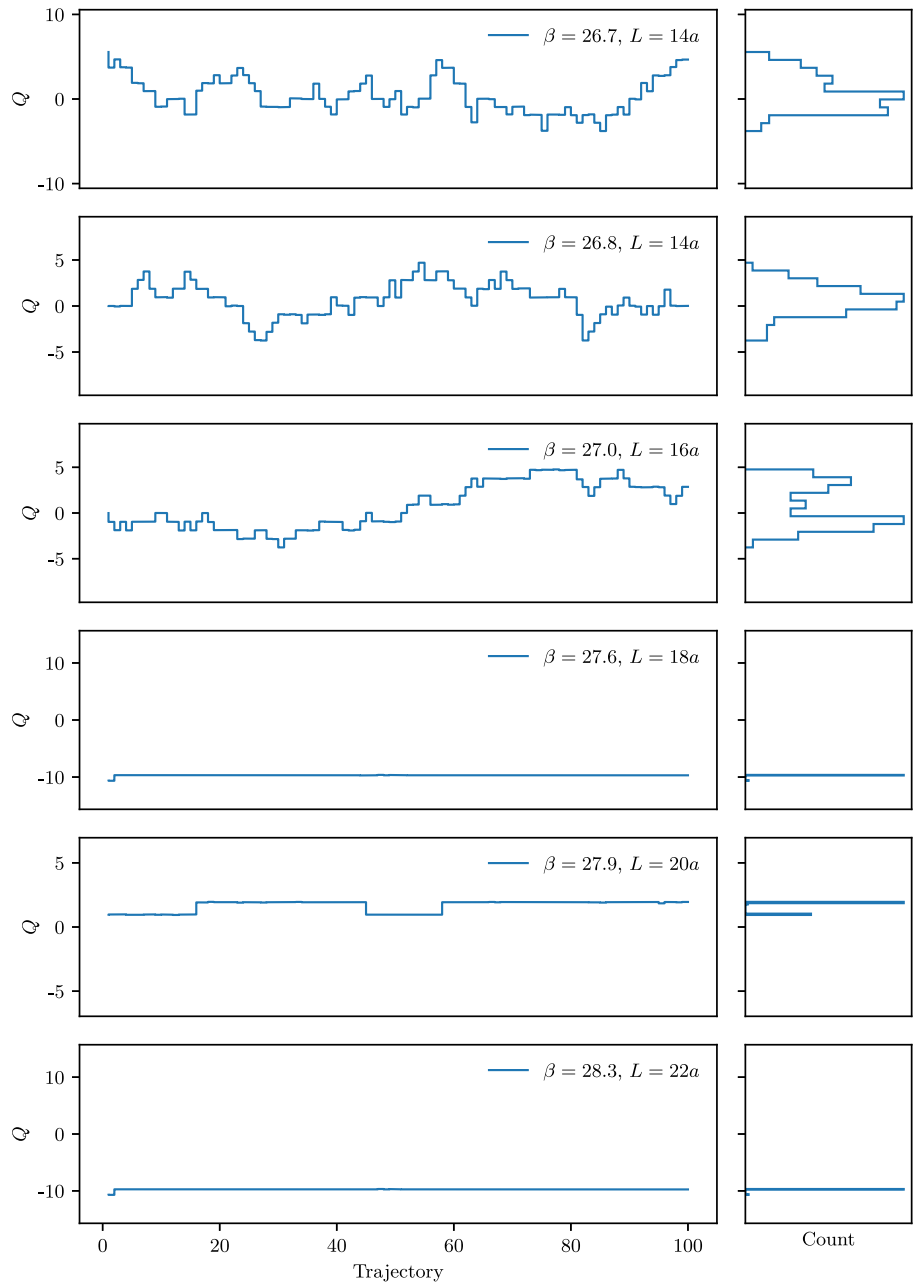

FIG. 15. Histories and statistical distributions (histograms) of the topological charge as defined in Eq. (11) for the ensembles obtained at $N = 4$. The configurations of each ensemble are smoothed with the gradient flow defined in Eq. (16). The frequency of sampling in running time is described in the text.

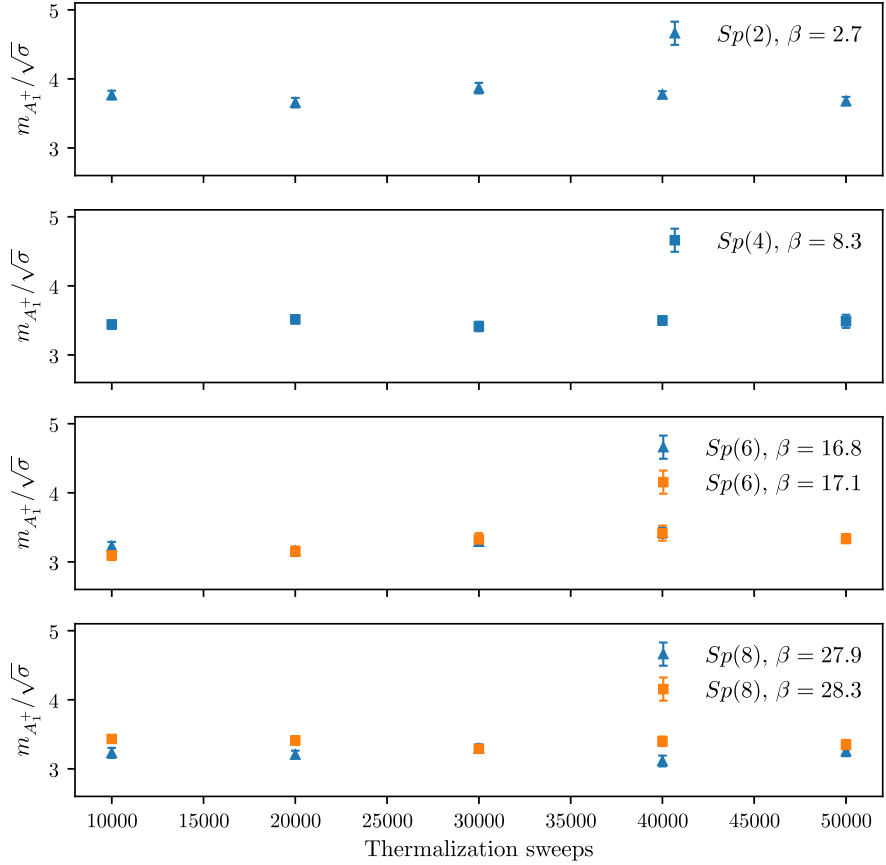

FIG. 16.   Mass of the $A_1^+$ glueball, in units of $\sqrt{\sigma}$, measured for different lengths of the initial thermalization. Each data point displays the running average over 10000 subsequent configurations, with the last one in the series having the sequential index corresponding to the abscissa.

$m_{R^P}/\sqrt{\sigma}$ in the continuum and $c_{R^P}$ as fitting parameters were performed for each symmetry channel. These linear fits are plotted as solid lines in Figs. 8–11, where the corresponding values of the $\chi^2/N_{\text{d.o.f.}}$ are also reported. Note that because of the way that the error on these measurements was evaluated, the $\chi^2/N_{\text{d.o.f.}}$ are slightly underestimated. These extrapolations are discussed in Sec. IV C.

## APPENDIX D: A CLOSER LOOK AT THE $Sp(4)$ DATA

In Sec. III we presented the spectrum of $Sp(2N)$ theories in the continuum and large-$N$ limits. Two sets of ensembles are available for $N = 2$: The one obtained for Ref. [7] (old ensembles) and another, independent one, obtained for the present work (new ensembles). The estimates shown in Table IV, in the column $N = 2$, are the weighted averages of the continuum limits obtained from the new and old ensembles. In this Appendix, we present separately the two analyses for the new and old ensembles for $N = 2$.

For the new ensembles, the continuum and large-$N$ extrapolated estimates can be found in Table XXI, in units of both $\sqrt{\sigma}$ (top) and $m_{E^+}$ (bottom). The former extrapolated values, together with the corresponding large-$N$ extrapolated results, are displayed, in units of $\sqrt{\sigma}$, in Fig. 9, and the large-$N$ extrapolation is shown in Fig. 17.

The old ensembles have been reanalyzed following the approach used in this work—see Sec. C 2. The results are displayed in Fig. 18. The continuum and large-$N$ extrapolated values can be found in Table XXII, in units of both $\sqrt{\sigma}$ (top) and $m_{E^+}$ (bottom), and are displayed, in units of $\sqrt{\sigma}$, in Figs. 19 and 20, respectively. In Fig. 21 the spectrum in the large-N limit is represented together with the finite-N one.

As expected, the estimates of the spectrum obtained from the two ensembles are statistically compatible. This justifies taking weighted averages as our best values for $N = 2$. We note that, while the new ensemble provides extrapolations with good values of $\chi^2/N_{\text{d.o.f.}}$, in the old ensemble higher values of the reduced $\chi^2$ are present. This hints toward slightly different systematics between the old and new simulations. This could explain the higher $\chi^2/N_{\text{d.o.f.}}$ for some extrapolations presented in the analysis in Sec. III. The broad compatibility of the data, nevertheless, suggests that the effect is not dominant.

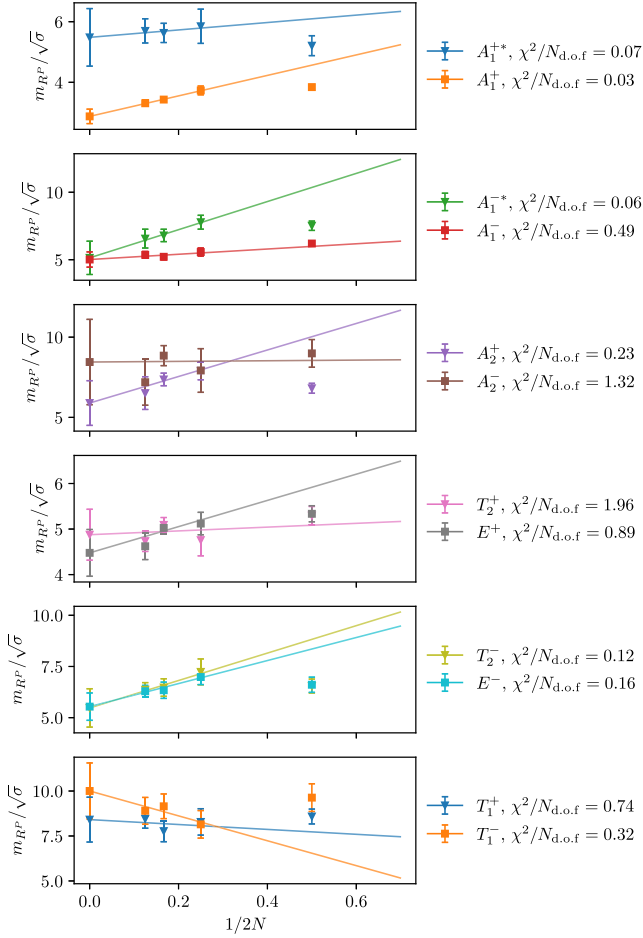

FIG. 17. Glueball mass in each symmetry channel $R^P$ of the $Sp(2N)$ theory, in units of $\sqrt{\sigma}a^2$, as a function of $1/2N$. For $N = 2$ only the numerical measurements reported in this publication were used (new ensembles). The point corresponding to $1/2N = 0$ is the value of $m_{R^P}/\sqrt{\sigma}(\infty)$ obtained from the best fit of Eq. (54) to the data. See main text for details.

TABLE XXI. In the left column, estimates of the spectrum at $N = 2$, in units of $\sqrt{\sigma}a$ and $m_{E^+}$. These are obtained from the data generated for this work (new ensembles). In the right column, the extrapolation to $N = \infty$ obtained from fits of Eq. (54) to the data using the data in the left column for $N = 2$ and the same data as before for $N = 1, 3, 4$.

| $R^P$ | $N = 2$ | | $\infty$ | |
|---|---|---|---|---|
| | $m_{R^P}/\sqrt{\sigma}$ | $\chi^2/N_{\text{d.o.f.}}$ | $m_{R^P}/\sqrt{\sigma}$ | $\chi^2/N_{\text{d.o.f.}}$ |
| $A_1^+$ | 3.73(15) | 0.69 | 3.209(91) | 1.15 |
| $A_1^{+*}$ | 5.85(56) | 0.04 | 5.89(37) | 0.14 |
| $A_1^-$ | 5.57(30) | 0.22 | 4.91(23) | 0.27 |
| $A_1^{-*}$ | 7.78(51) | 0.09 | 6.70(52) | 1.01 |
| $A_2^+$ | 7.90(56) | 0.89 | 7.70(52) | 1.1 |
| $A_2^-$ | 7.9(1.4) | 0.18 | 8.23(91) | 0.66 |
| $E^+$ | 5.12(25) | 0.71 | 4.77(19) | 0.64 |
| $E^-$ | 6.99(37) | 0.56 | 6.31(34) | 0.98 |
| $T_2^+$ | 4.76(35) | 0.82 | 4.78(21) | 1.0 |
| $T_2^-$ | 7.24(63) | 1.37 | 6.45(39) | 0.71 |
| $T_1^+$ | 8.28(73) | 0.43 | 8.02(55) | 0.43 |
| $T_1^-$ | 8.15(77) | 1.2 | 8.52(75) | 0.75 |
| | $N = 2$ | | $\infty$ | |
| | $m_{R^P}/m_{E^+}$ | $\chi^2/N_{\text{d.o.f.}}$ | $m_{R^P}/m_{E^+}$ | $\chi^2/N_{\text{d.o.f.}}$ |
| $A_1^+$ | 0.710(33) | 0.44 | 0.675(33) | 0.51 |
| $A_1^{+*}$ | 0.957(77) | 0.03 | 1.230(88) | 0.22 |
| $A_1^-$ | 1.159(54) | 0.18 | 1.005(69) | 0.58 |
| $A_1^{-*}$ | 1.40(10) | 0.34 | 1.50(13) | 0.28 |
| $A_2^+$ | 1.264(79) | 0.17 | 1.56(13) | 0.41 |
| $A_2^-$ | 1.66(18) | 0.13 | 1.76(20) | 0.12 |
| $E^-$ | 1.235(99) | 0.45 | 1.40(11) | 0.25 |
| $T_2^+$ | 0.968(56) | 0.13 | 1.037(58) | 0.18 |
| $T_2^-$ | 1.223(85) | 0.43 | 1.45(11) | 0.54 |
| $T_1^+$ | 1.59(11) | 0.14 | 1.68(14) | 1.58 |
| $T_1^-$ | 1.85(18) | 0.01 | 1.82(18) | 1.74 |

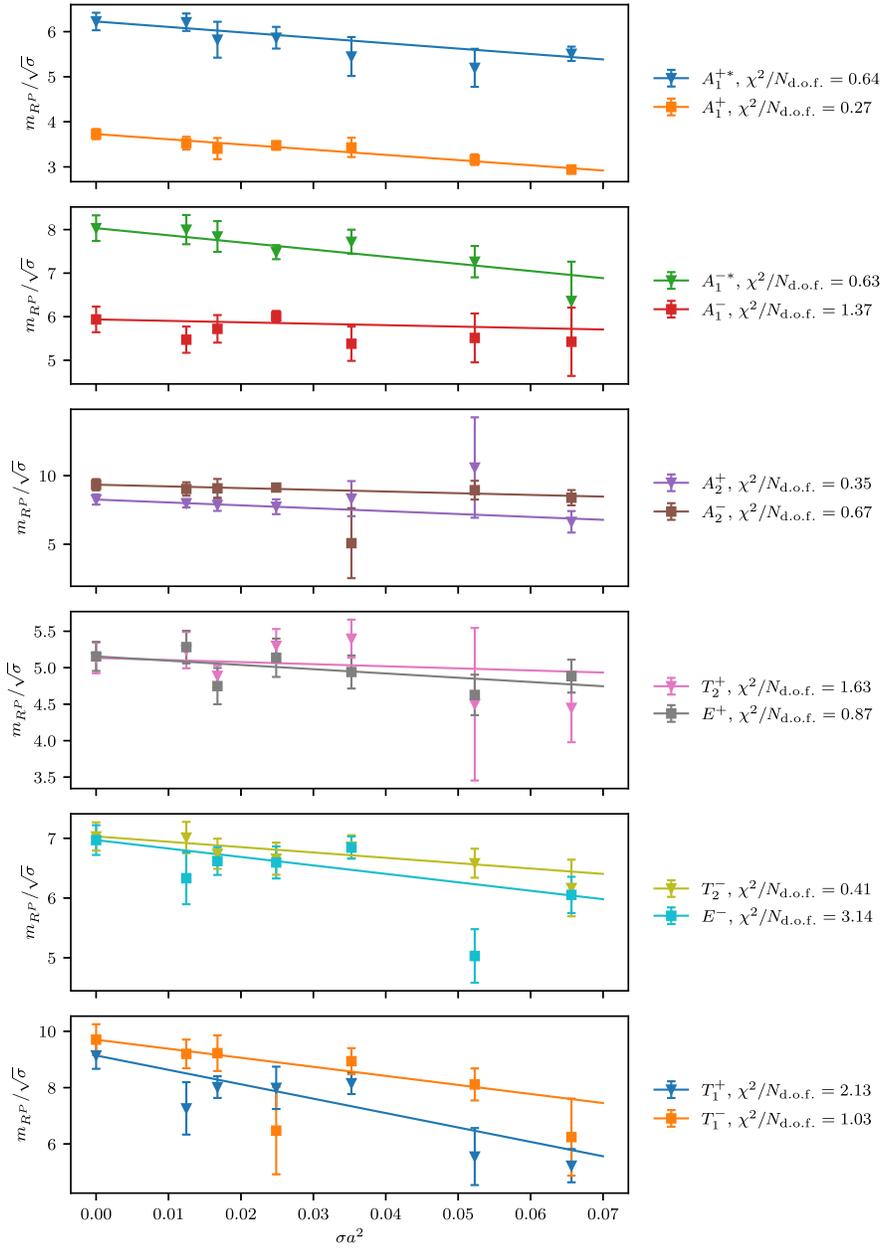

FIG. 18. Glueball mass in each symmetry channel $R^P$ of the $Sp(2N)$ theory with $N = 2$, in units of $\sqrt{\sigma}$, as a function of $\sigma a^2$. Only the data produced for Ref. [7] were used (old ensembles). For each symmetry channel $R^P$, the value at $\sigma a^2 = 0$ is the continuum limit, obtained from the best fit of Eq. (C1) to the data. The best fits lines are represented as solid lines. The extrapolated results are reported in Table XXII.

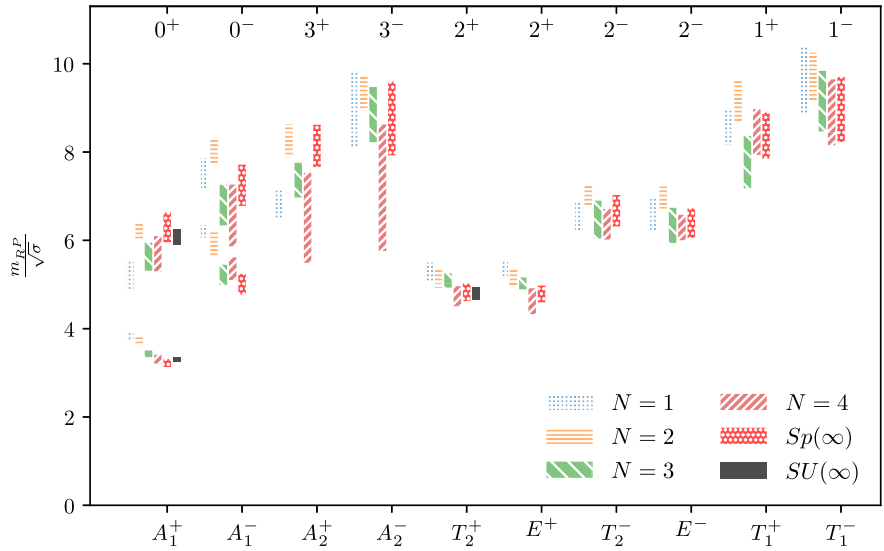

FIG. 19. Spectrum of the $Sp(2N)$ theory in the continuum limit for $N = 1, 2, 3, 4$ and $N = \infty$, in units of $\sqrt{\sigma}$ from the data collected for Ref. [7]. For ease of comparison, we have reported also the masses of the $A_1^{++}$ and $E^{++}$ channels for $SU(\infty)$ (borrowed from [2]).

TABLE XXII. Calculations of the masses in the continuum limit for $N = 2$ and each channel, in units of $\sqrt{\sigma}a$ and $m_{E^+}$, using only a reanalysis of the $N = 2$ data from Ref. [7] (old ensembles), as explained in the text.

| $R^P$ | $N = 2$ | | $\infty$ | |
|---|---|---|---|---|
| | $m_{R^P}/\sqrt{\sigma}$ | $\chi^2/N_{\text{d.o.f.}}$ | $m_{R^P}/\sqrt{\sigma}$ | $\chi^2/N_{\text{d.o.f.}}$ |
| $A_1^+$ | 3.73(11) | 0.27 | 3.222(90) | 1.65 |
| $A_1^{+*}$ | 6.23(19) | 0.64 | 6.29(34) | 3.1 |
| $A_1^-$ | 5.94(29) | 1.37 | 5.01(23) | 1.0 |
| $A_1^{-*}$ | 8.03(29) | 0.63 | 7.25(47) | 3.45 |
| $A_2^+$ | 8.26(37) | 0.35 | 8.14(48) | 3.28 |
| $A_2^-$ | 9.34(41) | 0.67 | 8.76(83) | 1.06 |
| $E^+$ | 5.16(20) | 0.87 | 4.79(19) | 0.71 |
| $E^-$ | 6.97(25) | 3.14 | 6.39(33) | 1.61 |
| $T_2^+$ | 5.13(21) | 1.63 | 4.83(20) | 0.69 |
| $T_2^-$ | 7.03(24) | 0.41 | 6.67(36) | 1.69 |
| $T_1^+$ | 9.14(47) | 2.13 | 8.37(52) | 1.58 |
| $T_1^-$ | 9.70(55) | 1.03 | 8.96(74) | 0.27 |

| $R^P$ | $N = 2$ | | $\infty$ | |
|---|---|---|---|---|
| | $m_{R^P}/m_{E^+}$ | $\chi^2/N_{\text{d.o.f.}}$ | $m_{R^P}/m_{E^+}$ | $\chi^2/N_{\text{d.o.f.}}$ |
| $A_1^+$ | 0.707(33) | 0.62 | 0.674(32) | 0.39 |
| $A_1^{+*}$ | 1.186(55) | 0.16 | 1.274(82) | 1.07 |
| $A_1^-$ | 1.092(73) | 0.82 | 1.010(69) | 0.59 |
| $A_1^{-*}$ | 1.554(85) | 0.89 | 1.56(13) | 0.77 |
| $A_2^+$ | 1.583(90) | 0.72 | 1.64(12) | 1.38 |
| $A_2^-$ | 1.76(11) | 1.0 | 1.80(19) | 0.09 |
| $E^-$ | 1.346(73) | 1.73 | 1.40(10) | 0.14 |
| $T_2^+$ | 1.040(58) | 0.57 | 1.050(57) | 0.23 |
| $T_2^-$ | 1.354(67) | 0.66 | 1.443(97) | 0.36 |
| $T_1^+$ | 1.83(12) | 1.71 | 1.78(14) | 2.38 |
| $T_1^-$ | 1.85(13) | 1.51 | 1.89(18) | 0.35 |

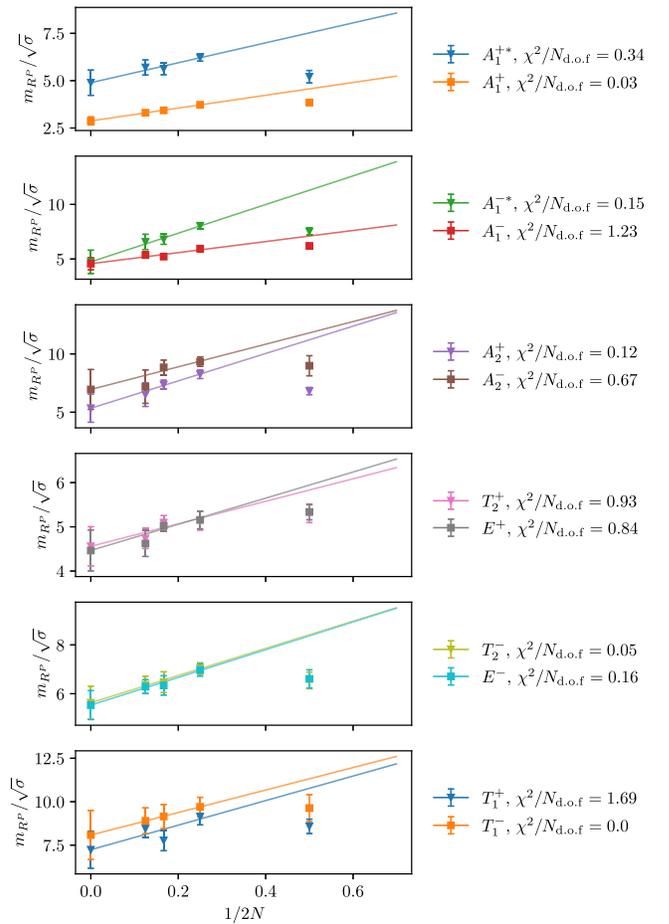

FIG. 20. Glueball mass in each symmetry channel $R^P$ in units of $\sqrt{\sigma}$, as a function of $1/2N$. For $N = 2$ only the data created for Ref. [7] were used. The point corresponding to $1/2N = 0$ is the value of $m_{R^P}/\sqrt{\sigma}(\infty)$ obtained from the best fit of Eq. (54) to the data.

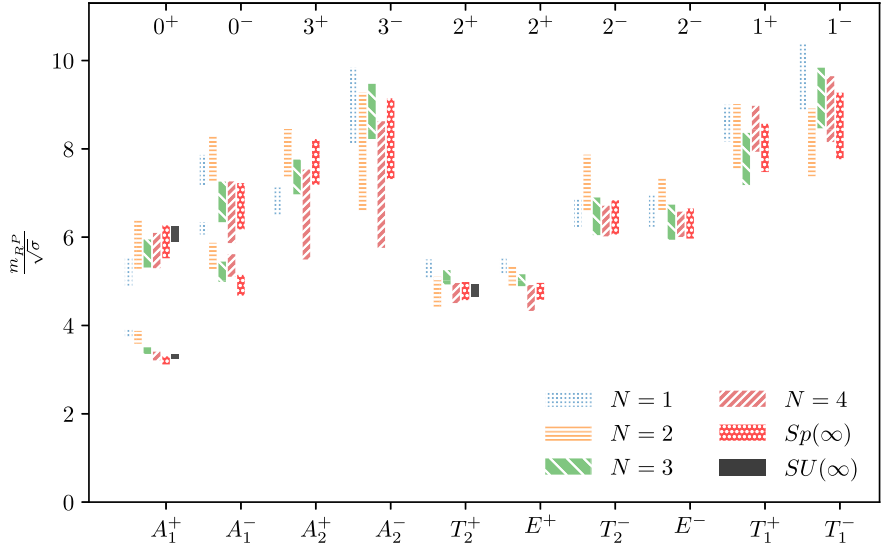

FIG. 21. Spectrum of the $Sp(2N)$ theory in the continuum limit for $N = 1, 2, 3, 4$ and $N = \infty$, in units of $\sqrt{\sigma}$. For $N = 2$, only the numerical measurements reported in this publication were used (new ensembles). For ease of comparison, we have reported also the masses of the $A_1^{++}$ and $E^{++}$ channels for $SU(\infty)$ (borrowed from [2]).

## APPENDIX E: ON THE INCLUSION OF $N = 1$ IN THE LARGE-$N$ EXTRAPOLATION

In Sec. III, the large-$N$ extrapolation of the spectrum has been provided including the value $N = 1$ for all channels. For a handful of channels, this gives a value of $\chi^2/N_{\text{d.o.f.}}$ above 2, indicative of a lower statistical significance of the extrapolation. This may suggest that, for these specific channels, the $N = 1$ value is not captured by the large-$N$ expansion. Indeed, in previous studies of $SU(N)$ gauge theories (e.g., Ref. [2]), the value of the $\chi^2/N_{\text{d.o.f.}}$ has been used as an indication of the reliability of the truncation of the large-$N$ series at a given order for capturing results at some finite value of $N$. In our work, excluding the data for $N = 1$ generally improves the value of the $\chi^2/N_{\text{d.o.f.}}$.

However, this leaves only three points for the extrapolation, and hence creates a larger systematic bias on the latter. Likewise, adding a higher-order correction will decrease the number of degrees of freedom and hence introduce more noise. Being faced with the necessity to make a choice, we have opted to systematically include $N = 1$ in all large-$N$ extrapolations. This means that we interpret a larger value of the $\chi^2/N_{\text{d.o.f.}}$ as results of fluctuations in the data or of some unknown systematics, rather than as stemming from the fact that $N = 1$ is not described by the expansion. The question is left open by this study. For completeness, we compare in Table XXIII our results for the extrapolations with $N = 1$ systematically included and excluded. Most of the results are compatible at the two sigma level.

TABLE XXIII. Large-$N$ extrapolated masses of the glueball spectrum obtained from a fit of Eq. (54), in the case in which the estimates at $N = 1$ are included (left) or excluded (right). Note that the left-hand part of this table is the same as the last column of Table IV and the same as Table V.

| $R^P$ | $m_{R^P}/\sqrt{\sigma}$ | $c_{R^P}$ | $\chi^2/N_{\text{d.o.f.}}$ | $m_{R^P}/\sqrt{\sigma}$ | $c_{R^P}$ | $\chi^2/N_{\text{d.o.f.}}$ |
|---|---|---|---|---|---|---|
| $A_1^+$ | 3.241(88) | 1.29(29) | 2.38 | 2.87(19) | 3.4(1.0) | 0.03 |
| $A_1^{+*}$ | 6.29(33) | $-1.6(1.2)$ | 2.91 | 4.94(66) | 4.9(3.0) | 0.3 |
| $A_1^-$ | 5.00(22) | 2.43(60) | 0.63 | 4.73(50) | 3.9(2.5) | 0.87 |
| $A_1^{-*}$ | 7.31(45) | 0.9(1.4) | 3.5 | 4.8(1.1) | 12.6(4.7) | 0.13 |
| $A_2^+$ | 8.22(46) | $-2.5(1.3)$ | 3.3 | 5.5(1.2) | 10.5(5.3) | 0.15 |
| $A_2^-$ | 8.69(83) | 1.3(3.0) | 0.9 | 7.2(1.7) | 8.4(7.5) | 0.73 |
| $T_2^+$ | 4.80(20) | 1.01(69) | 0.65 | 4.72(42) | 1.5(2.2) | 1.26 |
| $E^+$ | 4.79(19) | 1.15(63) | 0.72 | 4.52(42) | 2.6(2.1) | 0.9 |
| $T_2^-$ | 6.71(35) | 0.1(1.2) | 1.97 | 5.60(67) | 5.8(3.1) | 0.06 |
| $E^-$ | 6.44(33) | 0.9(1.2) | 2.03 | 5.52(57) | 5.8(2.7) | 0.16 |
| $T_1^+$ | 8.33(51) | 0.7(1.6) | 1.15 | 7.5(1.0) | 5.2(5.0) | 1.41 |
| $T_1^-$ | 8.76(72) | 1.7(2.6) | 0.02 | 8.8(1.3) | 1.7(6.3) | 0.03 |